\documentclass[aps,prb,
longbibliography, 
twocolumn, 
floatfix,
nofootinbib]{revtex4-2}
\usepackage{graphicx}
\usepackage{amsmath,amssymb,mathrsfs,amsthm}
\usepackage{svg}
\usepackage[pdftex, colorlinks=true, linkcolor=blue, citecolor=blue, urlcolor=blue]{hyperref}
\usepackage{bbold}
\usepackage{comment}
\usepackage{units}
\usepackage{times}

\usepackage{mathtools}
\usepackage{lipsum}
\usepackage{physics}

\usepackage{color}

\begin{document}

\title{Quantum theory of plasmon polaritons in chains of metallic nanoparticles:\\ From near- to far-field coupling regime}

\author{Thomas F.\ Allard}
\affiliation{Universit\'e de Strasbourg, CNRS, Institut de Physique et Chimie des Mat\'eriaux de Strasbourg, UMR 7504, F-67000 Strasbourg, France}
\author{Guillaume Weick}
\affiliation{Universit\'e de Strasbourg, CNRS, Institut de Physique et Chimie des Mat\'eriaux de Strasbourg, UMR 7504, F-67000 Strasbourg, France}

\begin{abstract}
We develop a quantum theory of plasmon polaritons in chains of metallic nanoparticles, describing both near- and far-field interparticle distances, by including plasmon--photon umklapp processes.
Taking into account the retardation effects of the long-range dipole--dipole interaction between the nanoparticles, which are induced by the coupling of the plasmonic degrees of freedom to the photonic continuum, we reveal the polaritonic nature of the normal modes of the system. 
We compute the dispersion relation and radiative linewidth, as well as the group velocities of the eigenmodes, and compare our numerical results to classical electrodynamic calculations within the point-dipole approximation. Interestingly, the group velocities of the polaritonic excitations present an almost periodic sign change and are found to be highly tunable by modifying the spacing between the nanoparticles.
We show that, away from the intersection of the plasmonic eigenfrequencies with the free photon dispersion, an analytical perturbative treatment of the light--matter interaction is in excellent agreement with our fully retarded numerical calculations.
We further study quantitatively the hybridization of light and matter excitations, through an analysis of Hopfield's coefficients. Finally, we consider the limit of infinitely spaced nanoparticles and discuss some recent results on single nanoparticles that can be found in the literature.
\end{abstract}

\maketitle

\section{Introduction}

Following the rise of miniaturization of physical systems, the issue of energy and information transport through light within structures with nanoscale dimensions has become increasingly important. 
Notably, a way to go beyond the diffraction limit encountered in usual dielectric waveguides, such as optical fibers, is highly sought-after, since it is expected to have applications in various ranges of modern physics, from integrated optical circuits or microscopy to biophotonics or data 
storage~\cite{Barnes2003, Stockman2011}.

An interesting proposal made more than 20 years ago is to use electrodynamic interparticle interactions in a linear chain of equidistant metallic nanoparticles to be utilized as an effective optical waveguide \cite{Quinten1998}. 
It is indeed well known that metallic nanoparticles present a strong absorption peak at the Mie frequency, due to the presence of a collective excitation, the localized surface plasmon (LSP), which corresponds to a dipolar collective oscillation of the conduction electrons of the nanoparticle \cite{Meier}. 
Systems based upon such collective excitations have been shown to achieve the confinement and control of light at the nanoscale \cite{Ozbay2006,Gramotnev_2010}. 

On the one hand, 
linear chains of nanoparticles
have been extensively studied in the past through numerous experimental \cite{Krenn1999,Maier2002,Maier2003,Koenderink2007,Crozier07,Apuzzo2013,Barrow2014} and theoretical \cite{Brongersma2000,Park,Weber,Citrin2004,Simovski2005,Citrin2006,Koenderink, Markel, Fung2007,Conforti2010,Udagedara,Rolly2012,Lee2012,Petrov2015,BrandstetterKunc,Downing2017,Compaijen,Pikalov} works, especially in the near-field regime where propagation is most promising, that is, when the interparticle distance $d$ is such that $d \ll \lambda_0$, $\lambda_0$ being the wavelength associated with the dipolar LSP excitation. On the other hand, the far-field regime, with $d \gtrsim \lambda_0$,  has been much less studied \cite{Pinchuck2008}, even if it also has  interesting properties. 
Notably, a topological phase transition was recently theoretically unveiled in the far-field regime of a dimerized plasmonic chain \cite{Pocock2019}, which represents a direct topological analog~\cite{Downing} of the regularly spaced chain studied here.

Both the near- and far-field regimes still remain challenging to realize experimentally since it requires cutting-edge technologies. 
Recently, however, impressive progress has been made in 
producing highly regular samples of near-field coupled gold nanoparticles  with long-range order \cite{Schulz2020,Mueller2020}.

After early theoretical studies focused on the quasistatic limit \cite{Quinten1998,Park,Brongersma2000}, the authors of Refs.~\cite{Citrin2004,Weber} noticed that retardation effects of the dipolar interaction along the chain are crucial and imply strong modifications of the energy spectrum. 
These works, followed by numerous others \cite{Citrin2006,Koenderink,Markel,Fung2007,Conforti2010,Udagedara,Petrov2015,Compaijen,Pikalov}, are based on the numerical resolution of the coupled-dipole equations in the point dipole approximation, which arise in the framework of Mie scattering theory. 
These are therefore fundamentally classical studies, which originate from Maxwell's equations.

In recent years, however, quantum nanoplasmonics has attracted a lot of attention in the community \cite{Tame2013}. 
The description of nanoscale optical properties of plasmonic metamaterials can exceed the limits of a classical description and requires a more general, quantum-mechanical treatment. 
For this purpose, approaches using quantum electrodynamics have been developed to tackle the properties of LSPs \cite{Miwa2021} and linear chains of metallic nanoparticles \cite{Lee2012,BrandstetterKunc}. 
A quantum approach turns out to be essential when the size of the particles are sufficiently small to present quantum-size effects, such as Landau damping \cite{Fernique2020}
or electron spill-out \cite{Weick2006,Downing2020}, namely, for radii smaller than approximately \unit[10--20]{nm}. 

Here, we build upon a Hamiltonian quantum theory of a chain of metallic nanoparticles \cite{BrandstetterKunc,Downing2017} which has the benefits to be an analytically tractable approach and to be able to readily incorporate quantum effects, such as those previously mentioned. It also permits the use of tools borrowed from quantum optics as well as nonrelativistic quantum electrodynamics, which are familiar to a growing part of the condensed matter community.

Several plasmonic systems have been studied with this quantum theory both in \cite{Mann2018,Downingb,Downing2021} and out \cite{BrandstetterKunca,Downinga,Downing2017,Lamowski} of a finite photonic cavity.
While for systems inside a cavity the plasmonic degrees of freedom couple mostly to a finite number of photonic modes, in vacuum the interaction with the whole continuum must be considered.
This makes the complete diagonalization of the Hamiltonian more difficult, especially for low-dimensional systems for which crystal momentum is not conserved for directions transverse to the plasmonic lattice, where translational symmetry is absent. 
For this reason, until now, a perturbative treatment of the light--matter interaction has been used for these 
low-dimensional systems, which does not take into account the polaritonic nature of the excitations but presents the advantage of remaining analytically tractable \cite{Downing2017}. 
However, the approach of Ref.~\cite{Downing2017} remains limited since it only considers the near-field regime. 
To extend the model of Ref.~\cite{Downing2017} to nanoparticles with interparticle distances in the far field, the effects of plasmon--photon umklapp processes must be included. To fully take into account the light--matter coupling, we rely on a diagonalization method elaborated by Hopfield \cite{Hopfield} in the context of exciton-polaritons. 
Such a method has already been employed for describing the plasmonic properties of 
three-dimensional lattices \cite{Lamowski}, and has been recently successfully compared with experimental 
results~\cite{Mueller2020}. 

The system which we study in the following consists of a one-dimensional chain of equidistant resonant dipolar meta-atoms interacting through a long-range retarded dipole--dipole interaction. 
A typical platform corresponding to our theory is spherical metallic nanoparticles hosting LSPs in the point-dipole approximation, but any other system where dipolar interactions are dominant could also be used, such as, e.g., plasmonic nanorods or microwave helical resonators \cite{Mann2018}.
In this paper, we focus on the consequences of the light--matter coupling on such a chain and hence on retardation effects of the Coulomb interaction. 

Here, we provide a full quantum description of the plasmon--polariton excitations propagating along the chain. 
Within our approach, we work in the Coulomb gauge such that the scalar and vector potentials describe, respectively, the longitudinal and transverse components of the electromagnetic field. 
Within this choice of gauge, the scalar potential accounts for purely instantaneous, long-range Coulomb interactions between the LSPs, leading to a collective plasmonic excitation extended over the whole array. 
The retardation effects arise from the light--matter coupling of the plasmonic modes with the transverse vector potential, which contains the photonic degrees of freedom, present in the vacuum electromagnetic field surrounding the chain. 
This leads to a hybridization of plasmons and transverse photons into plasmon--polaritons. 
Notably, the light--matter coupling results in a complex band structure which takes into account radiative losses.

The paper is organized as follows: Section~\ref{sec:Model} presents the  microscopic quantum model of the plasmonic chain coupled to vacuum electromagnetic modes. 
In Sec.~\ref{sec:Treatment of the Light Matter Coupling}, we describe how retardation effects are taken into account, either exactly with the numerical solution of a transcendental dispersion relation or perturbatively, using second-order perturbation theory to find approximate analytical expressions.
This allows us to provide both the band structure and the radiative decay rates, as well as the group velocities associated with the normal modes in Sec.~\ref{sec:Results}.
In the latter section, we also take advantage of our quantum formalism to examine the hybridization of light and matter degrees of freedom, and we study the limit of infinitely spaced nanoparticles.
Finally, we summarize our findings and draw conclusions in Sec.~\ref{sec:Conclusion}.
Four appendixes complement the discussion presented in the main text.

\section{Model}\label{sec:Model}

In this section, we present a microscopic Hamiltonian model of a chain of interacting spherical metallic nanoparticles which host LSPs in the point-dipole approximation. This approach relies on the quantum theory developed in Refs.~\cite{BrandstetterKunc, Downing2017} but with two substantial improvements: First, we consider here plasmon--photon umklapp scattering processes, which will allow us to study both the near- and far-field regimes of interaction. Second, we treat the retardation effects exactly and therefore reveal the polaritonic behavior of the collective excitations along the array. 
In Appendix~\ref{sec:AppendixClassicalModel}, we present a brief overview of the analogous, widely used macroscopic classical model based on the solution of the fully retarded Maxwell equations to compare it to our quantum approach.

\begin{figure}[tb]
 \includegraphics[width=\linewidth]{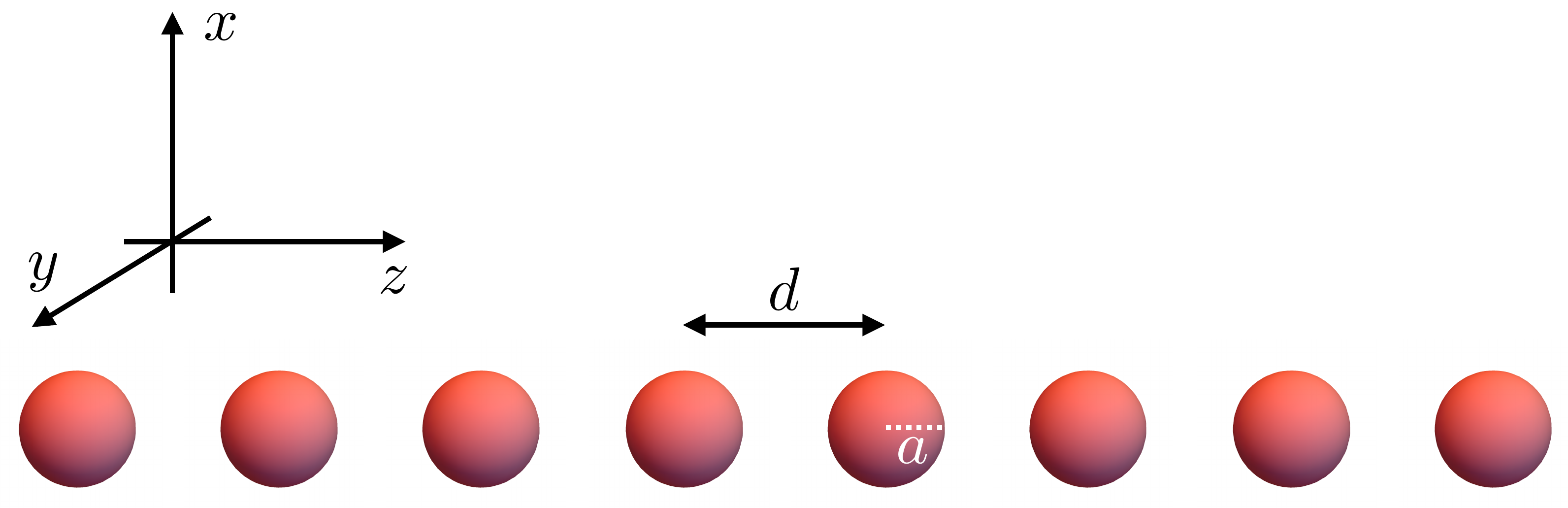}
 \caption{Sketch of a linear chain of identical spherical metallic nanoparticles of radius $a$, separated by a center-to-center distance $d$, arranged along the $z$ direction.}
 \label{fig:sketch}
\end{figure}

The chain, orientated along the $z$ direction, is embedded in vacuum and comprises $\mathcal{N} \gg 1$ nanoparticles\footnote{Such a limit has been shown to be a good approximation for chains with $\mathcal{N} \geqslant 20$ \cite{Weber,BrandstetterKunc}. } with radius $a$, separated by a center-to-center distance $d$ (see Fig.~\ref{fig:sketch}).
We consider the particular regime of nanoparticles with radius $a$ much smaller than the inverse wave vector $k_0^{-1}$ associated with the single LSP mode ($k_0a\ll1)$, such that each LSP can be considered as an oscillating point-dipole. 
We also consider center-to-center distances $d\gtrsim3a$. For such separations, one can neglect multipolar interactions \cite{Park} and tunneling effects between nanoparticles \cite{Scholl2013}.
Hence, each LSP corresponds to a harmonic dipole oscillation of the electronic center of mass at an individual resonance frequency $\omega_0=k_0 c$, with $c$ the speed of light in vacuum. 
For the case of alkaline nanoparticles in vacuum, $\omega_0 = \omega_{\mathrm{p}}/\sqrt{3}$, where $\omega_{\mathrm{p}}=\sqrt{4\pi n_{\mathrm{e}}e^2/m_{\mathrm{e}}}$ is the plasma frequency, $-e (<0)$ is the electron charge, $m_{\mathrm{e}}$ its mass, and $n_{\mathrm{e}}$ the electronic density.\footnote{Throughout this paper, we use cgs units.}

The LSPs in the chain couple through a long-range dipolar interaction, leading to collective plasmons extended over the whole array. The resulting band structure is then significantly modified by retardation effects. Indeed, due to the large size $(\mathcal{N}-1)d + 2a \simeq \mathcal{N}d$ of the chain, one must deal with a retarded dipole--dipole Coulomb interaction. In our quantum theory, the latter is encapsulated in the light--matter coupling of the plasmonic modes to a three-dimensional photonic environment \cite{Craig2012}.

Within our quantum formalism, quantum effects such as Landau damping \cite{Kawabata1966,Bertsch1994,Kreibig1995} and the associated shift it induces in the plasmonic resonance frequency \cite{Weick2006} can be readily incorporated in the theory by considering the interaction of the plasmon with electronic degrees of freedom (specifically, particle--hole excitations inside each nanoparticle). Landau damping has already been computed for a chain with nearest-neighbor interactions in Ref.~\cite{BrandstetterKunc}, while the associated electronic-induced redshift can be found in Ref.~\cite{ferni19_PhD}. Throughout this paper, however, we neglect the effects of an internal electronic environment and hence of Landau damping, since we are primarily focused on radiative effects, which are dominant as long as the nanoparticles are not too small (i.e., their radius should be larger than 10--\unit[20]{nm} \cite{Fernique2020}).

Using the Coulomb gauge \cite{CohenTannoudji1992,Craig2012}, the fully retarded Hamiltonian of the chain coupled to vacuum electromagnetic modes in a volume $\mathcal{V}=\mathcal{N}dL^2$ of linear size $L\xrightarrow[]{}\infty$ reads~\cite{Downing2017}
\begin{equation}
\label{eq:Hamiltonian}
H^{\sigma}=H^{\sigma}_{\mathrm{pl}} + H_\mathrm{ph} + H^{\sigma}_{\mathrm{pl}\mathrm{-}\mathrm{ph}},
\end{equation}
where $\sigma$ is a fixed parameter accounting for either one of the transverse $(\sigma=x,y)$ or the longitudinal $(\sigma=z)$ polarization of the plasmonic modes, according to the chosen experimental configuration. 
We emphasize here the fact that the volume $\mathcal{V}$ of the photonic cavity is infinite, the photonic environment representing a continuum of states, in contrast with situations with a finite cavity which could induce strong  coupling \cite{Downing2021}. We also note that, recently, techniques have been developed to quantize the electromagnetic field not through the Fourier components of the vector potential as we shall be doing here, but directly in position space \cite{Southall2021, Hodgson2021}.

The purely plasmonic Hamiltonian describing the LSPs coupled through the long-ranged quasistatic dipole--dipole interaction is
\begin{align}
    \label{eq:H_pl}
  H^{\sigma}_{\mathrm{pl}} =&\, \hbar \omega_0 \sum_{n=1}^\mathcal{N} {b_{n}^{\sigma}}^{\dagger} b_{n}^{\sigma} \nonumber
  \\
&\,+ \frac{\hbar \Omega}{2} \sum_{\substack{n,m=1\\(n\neq m)}}^\mathcal{N} \frac{\eta^{\sigma}}{|n-m|^3} \left( b_{n}^{\sigma} + {b_{n}^{\sigma}}^{\dagger} \right) \left( b_{m}^{\sigma} + {b_{m}^{\sigma}}^{\dagger} \right).
\end{align}
Here, the bosonic operator $b_{n}^{\sigma}$ (${b_{n}^{\sigma}}^{\dagger}$) annihilates (creates) an LSP with polarization $\sigma$ on nanoparticle $n \in [1,\mathcal{N}]$. The long-range dipolar coupling strength between LSPs scales as the inverse interparticle distance cubed and is given by 
\begin{equation}
\label{eq:Omega}
    \Omega = \frac{\omega_0}{2}\left(\frac{a}{d}\right)^3,
\end{equation}
and the polarization-dependent factor $\eta^{x,y} = 1$ ($\eta^{z} = -2$) for the transverse (longitudinal) mode accounts for the anisotropy of the quasistatic dipole--dipole interaction.

The photonic environment in Eq.~\eqref{eq:Hamiltonian} is described by the Hamiltonian
\begin{equation}
\label{eq:H_ph}
 H_{\mathrm{ph}} = \sum_l \sum_{\mathbf{k}, \hat{\lambda}^l_{\mathbf{k}}} \hbar \nu^l_{\mathbf{k}} {c_{\mathbf{k}}^{l,\hat{\lambda}^l_{\mathbf{k}}}}^{\dagger} c_{\mathbf{k}}^{l,\hat{\lambda}^l_{\mathbf{k}}}, 
\end{equation}
where the upper index $l$ on $\mathbf{k}$-dependent variables is a short form signifying that the quantity depends on $\mathbf{k}-\mathbf{G}^l$, with $\mathbf{G}^l = 2\pi l\hat{z}/d$ ($l \in \mathbb{Z}$) representing the set of reciprocal lattice vectors.
Hence the ladder operator $c_{\mathbf{k}}^{l,\hat{\lambda}^l_{\mathbf{k}}}$ (${c_{\mathbf{k}}^{l,\hat{\lambda}^l_{\mathbf{k}}}}^{\dagger}$) annihilates (creates) a photon with wave vector $\mathbf{k} - \mathbf{G}^l$ in the first Brillouin zone, transverse polarization $\hat{\lambda}^l_{\mathbf{k}}=\hat{\lambda}_{\mathbf{k}-\mathbf{G}^l}$, i.e., $(\mathbf{k}-\mathbf{G}^l)\cdot\hat{\lambda}_{\mathbf{k}-\mathbf{G}^l}=0$, and angular frequency 
\begin{equation}
    \nu^l_{\mathbf{k}} = c |\mathbf{k}-\mathbf{G}^l|,
\end{equation}
where hats designate unit vectors.
The two photon polarizations are parameterized as $\hat{\lambda}^{l,(1)}_{\mathbf{k}} = [\hat{z} \times (\mathbf{k}-\mathbf{G}^l)]/|\hat{z} \times (\mathbf{k}-\mathbf{G}^l)|$ and $\hat{\lambda}^{l,(2)}_{\mathbf{k}} = [(\mathbf{k}-\mathbf{G}^l) \times \hat{\lambda}^{l,(1)}_{\mathbf{k}}]/|(\mathbf{k}-\mathbf{G}^l) \times \hat{\lambda}^{l,(1)}_{\mathbf{k}}|$ for the case $\mathbf{k}-\mathbf{G}^l \neq \hat{z}$, while we choose $\hat{\lambda}^{l,(1)}_{\mathbf{k}} = \hat{x}$ and $\hat{\lambda}^{l,(2)}_{\mathbf{k}} = \hat{y}$ for the case $\mathbf{k}-\mathbf{G}^l = \hat{z}$.
We note that in previous works \cite{BrandstetterKunc,Downing2017}, only the photonic band $l = 0$ was considered.
This amounts to neglect umklapp processes and limits the theory to $k_0d \ll 1$, i.e., to short center-to-center distances $d$. Taking into account umklapp processes is therefore essential to describe interacting nanoparticles in the far-field region.

The plasmon--photon coupling Hamiltonian in Eq.~\eqref{eq:Hamiltonian} reads in the long-wavelength point-dipole approximation ($|\mathbf{k}|a\ll1$) \cite{Craig2012}
\begin{equation}
\label{eq:H_plph}
 H^{\sigma}_{\mathrm{pl}\textrm{-}\mathrm{ph}} = \frac{e}{m_{\mathrm{e}}c} \sum_{n=1}^{\mathcal{N}} \mathbf{\Pi}^{\sigma}_n \cdot \mathbf{A} (\mathbf{d}_n), 
\end{equation}
where $\mathbf{d}_n = d (n-1) \hat{z}$ corresponds to the location of the center of nanoparticle $n$. The momentum associated with the LSP in nanoparticle $n$ reads
\begin{equation}
\mathbf{\Pi}^{\sigma}_n=\mathrm{i}\sqrt{\frac{N_\mathrm{e}m_\mathrm{e}\hbar\omega_0}{2}}\left({b_n^\sigma}^\dagger-b_n^\sigma\right)\hat\sigma,
\end{equation}
where $N_{\mathrm{e}}$ is the number of electrons in a given nanoparticle, while the quantized vector potential is given by
\begin{align}
\label{eq:A}
\mathbf{A}(\mathbf{d}_n)=&\,\sum_l\sum_{\mathbf{k}, \hat\lambda^l_{\mathbf{k}}}
\hat\lambda^l_{\mathbf{k}}\sqrt{\frac{2\pi\hbar c^2}{\mathcal{V}\nu^l_{\mathbf{k}}}} \nonumber\\
&\, \times\left(
c_{\mathbf{k}}^{l,\hat\lambda^l_{\mathbf{k}}}\mathrm{e}^{\mathrm{i}(\mathbf{k}-\mathbf{G}^l)\cdot \mathbf{d}_n} + \mathrm{H.c.}
\right)\Theta\left(\omega_{\mathrm{c}} - \nu^l_{\mathbf{k}} \right).
\end{align}
In the equation above, we have introduced a Heaviside step function $\Theta(x)$ to take into account the finite size of the nanoparticles and to switch off the light--matter interaction for photonic modes with a frequency $\nu_{\mathbf{k}}^l$ larger than the cutoff frequency $\omega_{\mathrm{c}}=c/a$, for which the point-dipole approximation breaks down.\footnote{In our situation, the point-dipole approximation amounts to considering that exponentials of the type $\mathrm{e}^{\mathrm{i}\mathbf{k}\cdot\mathbf{r}}$ equal unity, where $\mathbf{r}$ is the position of the electrons in the nanoparticle.
Since at most $r\simeq a$, we require that $k\ll1/a$.
Therefore, we consider that the point-dipole approximation is accurate up to the wave number $k_{\mathrm{c}}=\Lambda/a$, where $\Lambda\lesssim1$ is a cutoff parameter which must be close to unity. In what follows, we set $\Lambda=1$.}

A significant simplification we are making in writing the coupling Hamiltonian \eqref{eq:H_plph} is to disregard the quadratic $A^2$ term, known as the diamagnetic term, which accounts for a photon self-interaction energy and does not contribute to retardation effects.
In the light--matter coupling regime considered in this paper, the diamagnetic term can be safely neglected, as it would only lead to very slight changes to the results presented here.
Details about the $A^2$ term are given in Appendix~\ref{sec:AppendixDiamagneticTerm}, as well as justifications of the above-mentioned simplification.

\section{Exact and perturbative treatments of the light--matter coupling} 
\label{sec:Treatment of the Light Matter Coupling}
\subsection{Full diagonalization of the polaritonic Hamiltonian}
\label{sec:Light Matter Interaction Hopfield}

The light--matter coupling Hamiltonian \eqref{eq:H_plph} includes the retardation effects of the dipole--dipole interaction along the chain, which modify the nature of the normal modes which become plasmon--polaritons.
Taking such a coupling into account is therefore essential and will be the subject of this section.

Considering the infinite chain limit with $\mathcal{N}\gg1$, it is convenient to choose, without loss of generality, periodic boundary conditions. 
We then move into wave vector space using the Fourier transform $b_n^{\sigma} = \mathcal{N}^{-1/2}\sum_q\mathrm{e}^{\mathrm{i}nqd}b_q^{\sigma}$, where the plasmonic wave number is $q = 2\pi p/\mathcal{N}d$, with the integer $p \in [-\mathcal{N}/2, +\mathcal{N}/2]$.
It is important to note that the periodicity of the chain in the $z$ direction implies that the longitudinal component of the photonic wave vector $\mathbf{k}$ is conserved with the plasmonic one.
We thus choose to write $k_z=q$ and, for commodity, we denote the $\mathbf{k}$-dependency as $\mathbf{k} = (\boldsymbol{\kappa},q)$ where $\boldsymbol{\kappa}=(k_x,k_y)$.

The Hamiltonian $\eqref{eq:Hamiltonian}$ can hence be rewritten in Fourier space as $H^{\sigma} = \sum_q H^{\sigma}_q$, where the $q$-dependent polaritonic Hamiltonian is
\begin{equation}
    \label{eq:fourier polaritonic Hamiltonian}
    H^{\sigma}_q = H^{\sigma}_{{\mathrm{pl}},q} + H_{\mathrm{ph},q} + H^{\sigma}_{\mathrm{pl-ph},q}.
\end{equation}
The plasmonic part can be written as
\begin{equation}
\label{eq:fourier plasmonic Hamiltonian}
    H^{\sigma}_{\mathrm{pl},q} = \hbar\omega_0b_q^{\sigma\dagger}b_q^{\sigma} + \frac{\hbar\Omega}{2}\left[f_q^{\sigma}b_q^{\sigma\dagger}(b_q^{\sigma} + b_{-q}^{\sigma\dagger}) + \mathrm{H.c.}\right],
\end{equation}
the photonic one as
\begin{equation}
\label{eq:fourier photonic Hamiltonian}
    H_{\mathrm{ph},q} = \sum_l\sum_{\boldsymbol{\kappa},\hat{\lambda}^l_{\boldsymbol{\kappa} q}}\hbar\nu^l_{\boldsymbol{\kappa} q}c^{l,\hat{\lambda}^l_{\boldsymbol{\kappa} q}\dagger}_{\boldsymbol{\kappa} q}c^{l,\hat{\lambda}^l_{\boldsymbol{\kappa} q}}_{\boldsymbol{\kappa} q},
\end{equation}
and the coupling as
\begin{align}
\label{eq:fourier coupling Hamiltonian}
    H^{\sigma}_{\mathrm{pl-ph},q} = &\, \mathrm{i}\hbar\omega_0\sum_l\sum_{\boldsymbol{\kappa},\hat{\lambda}^l_{\boldsymbol{\kappa} q}} \xi^l_{\boldsymbol{\kappa} q}(\hat{\sigma}\cdot\hat{\lambda}^l_{\boldsymbol{\kappa} q}) \nonumber \\
    &\,\times\left[\mathrm{e}^{-\mathrm{i}qd}c^{l,\hat{\lambda}^l_{\boldsymbol{\kappa} q}}_{\boldsymbol{\kappa} q}(b_q^{\sigma\dagger} - b_{-q}^{\sigma}) - \mathrm{H.c.}\right].
\end{align}
The lattice sum $f_q^{\sigma}$ in Eq.~\eqref{eq:fourier plasmonic Hamiltonian} can be expressed in closed form in terms of the polylogarithm function $\text{Li}_s(z) = \sum_{n=1}^{\infty}z^n/n^s$ as
\begin{equation}
 f_q^{\sigma} = \eta^{\sigma}\left[\mathrm{Li}_3(\mathrm{e}^{\mathrm{i}qd}) + \mathrm{Li}_3(\mathrm{e}^{-\mathrm{i}qd})\right],
\end{equation}
and the light--matter coupling strength in Eq.~\eqref{eq:fourier coupling Hamiltonian} is encapsulated in
\begin{equation}
    \label{eq:light matter coupling}
   \xi^l_{\boldsymbol{\kappa} q} = \sqrt{\dfrac{\omega_0\pi a^3}{\nu^l_{\boldsymbol{\kappa} q}dL^2}}\Theta\left(\omega_\mathrm{c} - \nu^l_{\boldsymbol{\kappa}q} \right).
\end{equation}

We start by diagonalizing exactly the Hamiltonian \eqref{eq:fourier polaritonic Hamiltonian}. Toward this goal, we define the bosonic plasmon--polariton annihilation operator $\mu^{\sigma}_q$ as a linear combination of plasmon and photon ladder operators,
\begin{align}
\label{eq:polaritonic operator}
    \mu^{\sigma}_q =&\, W_q^{\sigma}b_q^{\sigma} + X_q^{\sigma}b_{-q}^{\sigma\dagger} \nonumber\\
    &\,+ \sum_l\sum_{\boldsymbol{\kappa},\hat\lambda^l_{\boldsymbol{\kappa} q}} \left(Y_{\boldsymbol{\kappa} q}^{l,\sigma,\hat\lambda^l_{\boldsymbol{\kappa} q}}c_{\boldsymbol{\kappa} q}^{l,\hat{\lambda}^l_{\boldsymbol{\kappa} q}} + Z_{\boldsymbol{\kappa} q}^{l,\sigma,\hat\lambda^l_{\boldsymbol{\kappa} q}}c_{\boldsymbol{\kappa},
    -q}^{l,\hat{\lambda}^l_{\boldsymbol{\kappa}, -q}\dagger}\right),
\end{align}
where the Hopfield coefficients $W_{q}^{\sigma}$, $X_{q}^{\sigma}$, $Y_{\boldsymbol{\kappa} q}^{l,\sigma,\hat\lambda^l_{\boldsymbol{\kappa} q}}$, and $Z_{\boldsymbol{\kappa} q}^{l,\sigma\hat\lambda^l_{\boldsymbol{\kappa} q}}$ are 
\textit{a priori} complex. Due to the bosonic nature of plasmon--polaritons, the coefficients are normalized following the condition
\begin{equation}
\label{eq:normalization}
    \left|W_{q}^{\sigma}\right|^2 - \left|X_{q}^{\sigma}\right|^2 + \sum_l\sum_{\boldsymbol{\kappa},\hat\lambda^l_{\boldsymbol{\kappa} q}} \left( \left|Y_{\boldsymbol{\kappa} q}^{l,\sigma,\hat\lambda^l_{\boldsymbol{\kappa} q}}\right|^2 - \left|Z_{\boldsymbol{\kappa} q}^{l,\sigma,\hat\lambda^l_{\boldsymbol{\kappa} q}}\right|^2\right) = 1.
\end{equation}
We note that in the one-dimensional geometry which we consider here, the polariton polarization is aligned with the plasmonic one, $\sigma$. In higher-dimensional metamaterials, this is generally not the case \cite{Lamowski,Fernique2020}.

We now use the ansatz
\begin{equation}
\label{eq:ansatz}
    \mu^{\sigma}_q (t) = \mu^{\sigma}_q\,\mathrm{e}^{-\mathrm{i}\Omega^{\sigma}_{q}t},
\end{equation}
where, in the following, $\Omega^{\sigma}_q$ are referred to as the polaritonic eigenfrequencies of the system, to solve the Heisenberg equation of motion 
\begin{equation}
\label{eq:heisenberg equation}
    \dot{\mu}^{\sigma}_q(t) = \frac{\mathrm{i}}{\hbar}[H^{\sigma}, \mu^{\sigma}_q(t)],
\end{equation}
that we can rewrite as $\hbar\Omega^{\sigma}_q\mu^{\sigma}_q = [\mu^{\sigma}_q , H^{\sigma}]$. 
We then compute the above commutator using Eq.~\eqref{eq:polaritonic operator} and the bosonic commutation relations for the plasmonic and photonic ladder operators, respectively, $[b_q^{\sigma}, b_{q'}^{\sigma'\dagger}] = \delta_{q,q'}\delta_{\sigma,\sigma'}$ and $[c_{\boldsymbol{\kappa}q}^{l,\hat\lambda^{l,(i)}_{\boldsymbol{\kappa}q}}, c_{\boldsymbol{\kappa}'q'}^{l',\hat\lambda^{l',(j)}_{\boldsymbol{\kappa}'q'}\dagger}] = \delta_{q,q'}\delta_{\boldsymbol{\kappa}\boldsymbol{\kappa}'}\delta_{l,l'}\delta_{i,j}$. 
Since the ladder operators are independent from each other, we use them to factorize the equation and separate it in a system of four coupled equations with five unknown variables: the four Hopfield coefficients and the desired eigenfrequency $\Omega^{\sigma}_q$ for a given wave number $q$,
\begin{widetext}
\begin{subequations}
\label{eq:system of equations}
\begin{align}
    W_{q}^{\sigma} &= -\frac{1}{\Omega_q^{\sigma} - \omega_0 - \Omega f_q^{\sigma}}\left\{ \Omega f_q^{\sigma}X_{q}^{\sigma} 
    +\mathrm{i}\omega_0\mathrm{e}^{\mathrm{i}qd}\sum_l\sum_{\boldsymbol{\kappa},{\hat\lambda^l_{\mathbf{\boldsymbol{\kappa}q}}}} \left[\xi^l_{\boldsymbol{\kappa} q}(\hat{\sigma}\cdot\hat\lambda^l_{\boldsymbol{\kappa} q})Y_{\boldsymbol{\kappa} q}^{l,\sigma,\hat\lambda^l_{\boldsymbol{\kappa} q}} 
    - \xi^l_{\boldsymbol{\kappa},-q}(\hat{\sigma}\cdot\hat\lambda^l_{\boldsymbol{\kappa},-q})Z_{\boldsymbol{\kappa} q}^{l,\sigma,\hat\lambda^l_{\boldsymbol{\kappa} q}}  \right] \right\},
    \\
    X_{q}^{\sigma} &=  \frac{1}{\Omega_q^{\sigma} + \omega_0 + \Omega f_q^{\sigma}}\left\{ \Omega f_q^{\sigma}W_{q}^{\sigma}
    \mathrm{i}\omega_0\mathrm{e}^{\mathrm{i}qd}\sum_l\sum_{\boldsymbol{\kappa},{\hat\lambda^l_{\mathbf{\boldsymbol{\kappa}q}}}} \left[\xi^l_{\boldsymbol{\kappa} q}(\hat{\sigma}\cdot\hat\lambda^l_{\boldsymbol{\kappa} q})Y_{\boldsymbol{\kappa} q}^{l,\sigma,\hat\lambda^l_{\boldsymbol{\kappa} q}}
    - \xi^l_{\boldsymbol{\kappa},-q}(\hat{\sigma}\cdot\hat\lambda^l_{\boldsymbol{\kappa},-q})Z_{\boldsymbol{\kappa} q}^{l,\sigma,\hat\lambda^l_{\boldsymbol{\kappa} q}}  \right]\right\}, 
\\
    Y_{\boldsymbol{\kappa} q}^{l,\sigma,\hat\lambda^l_{\boldsymbol{\kappa} q}} &= \mathrm{i}\omega_0\frac{\mathrm{e}^{-\mathrm{i}qd}\xi^l_{\boldsymbol{\kappa} q}(\hat{\sigma}\cdot\hat\lambda^l_{\boldsymbol{\kappa} q})}{ \Omega_q^{\sigma} - \nu^l_{\boldsymbol{\kappa}q} }\left(W_{q}^{\sigma} + X_{q}^{\sigma}\right),
\\
    Z_{\boldsymbol{\kappa} q}^{l,\sigma,\hat\lambda^l_{\boldsymbol{\kappa} q}}  &= \mathrm{i}\omega_0\frac{\mathrm{e}^{-\mathrm{i}qd}\xi^l_{\boldsymbol{\kappa},-q}(\hat{\sigma}\cdot\hat\lambda^l_{\boldsymbol{\kappa},-q})}{\Omega_q^{\sigma} + \nu^l_{\boldsymbol{\kappa},-q}}\left(W_{q}^{\sigma} + X_{q}^{\sigma}\right).
\end{align}
\end{subequations}
\end{widetext}
Solving the system of equations \eqref{eq:system of equations} for $\Omega_q^{\sigma}$ leads to the following transcendental equation,
\begin{equation}
\label{eq:equation_eigenfreq_sum}
    \frac{  {\Omega^{\sigma}_q}^2 - {\omega_q^{\sigma}}^2 }{ 4\omega_0{\omega_q^{\sigma}}^2 }  =  \sum_l \sum_{\boldsymbol{\kappa},{\hat\lambda^l_{\mathbf{\boldsymbol{\kappa}q}}}}\frac{\nu_{\boldsymbol{\kappa}q}^l(\hat{\sigma}\cdot\hat\lambda^l_{\mathbf{\boldsymbol{\kappa}q}})^2\left(\xi^l_{\boldsymbol{\kappa}q}\right)^2}{{\Omega^{\sigma}_q}^2 - {\nu^l_{\boldsymbol{\kappa}q}}^2},
\end{equation}
where the summation over $\boldsymbol{\kappa}$ excludes singular terms. 
Note that $\Omega^{\sigma}_q$ is \textit{a priori} complex, its imaginary part taking into account the radiative decay of the collective excitation inside the three-dimensional photonic continuum, leading to a radiative contribution to the linewidth [cf.\ Eq.~\eqref{eq:ansatz}].

Setting the light--matter coupling \eqref{eq:light matter coupling} to zero in Eq.~\eqref{eq:equation_eigenfreq_sum} naturally leads to $\Omega_q^{\sigma} = \omega_q^{\sigma}$, where 
\begin{equation}
\label{eq:quasistatic}
\omega_q^{\sigma} = \omega_0\sqrt{1+2\frac{\Omega}{\omega_0}f_q^{\sigma}}
\end{equation}
is the quasistatic spectrum of the collective plasmonic modes.
Such a quasistatic dispersion can also be obtained \cite{Downing2017} through a direct diagonalization of the Hamiltonian \eqref{eq:H_pl} by means of a bosonic Bogoliubov transformation (see Sec.~\ref{sec:Light Matter Interaction Perturbative}).

In Eq.~\eqref{eq:equation_eigenfreq_sum}, we can carry out the summation over photon polarizations using the relation 
\begin{equation}
\label{eq:polarization sum}
    \sum_{\hat{\lambda}^l_{\boldsymbol{\kappa} q}}\left(\hat{\sigma}\cdot\hat{\lambda}^l_{\boldsymbol{\kappa} q}\right)^2 = 1 - \left(\hat{\sigma}\cdot\frac{\mathbf{k}-\mathbf{G}^l}{\left|\mathbf{k}-\mathbf{G}^l\right|}\right)^2.
\end{equation}
To then compute the summation over $\boldsymbol{\kappa}$, we recall that we consider a quantization volume $\mathcal{V}\xrightarrow[]{} \infty$ such that we can take the continuum limit
\begin{equation}
   \frac{(2\pi)^2}{L^2}\sum_{\boldsymbol{\kappa}} \xrightarrow\,  \mathcal{P}\int_0^{\infty}\kappa\mathrm{d}\kappa\int_0^{2\pi} \mathrm{d}\theta,
\label{eq:continuum limit}
\end{equation}
where we used polar coordinates and $\mathcal{P}$ represents a principal value. The angular integral is trivial, however, in the radial one, the integrand suffers from an ultraviolet logarithmic divergence. 
The latter is naturally regularized with the help of the UV frequency cutoff $\omega_\mathrm{c}$ appearing in the light--matter coupling \eqref{eq:light matter coupling} and which takes into account the finite size of the nanoparticles.
The cutoff also selects how many photonic bands $l$ interact with the plasmonic chain, leaving us with a finite summation over the index $l$, from $-l_{\mathrm{max}}$ to $+l_{\mathrm{max}}$. In the first Brillouin zone, one can show that using a cutoff $\omega_\mathrm{c}=c/a$ leads to
    \begin{equation}
        l_{\mathrm{max}} = \left\lfloor\frac{d}{2\pi a}+\frac{1}{2}\right\rfloor,
    \end{equation}
where $\lfloor x \rfloor$ denotes the floor function.
After a long but straightforward calculation, Eq.~\eqref{eq:equation_eigenfreq_sum} translates into
\begin{align}
        \label{eq:Eigenfreq_exact_developed}
        {\Omega^{\sigma}_q}^2 - {\omega_q^{\sigma}}^2 =&\, \eta^{\sigma}\omega_0^2{\omega_q^{\sigma}}^2\frac{a^3}{dc^2}\sum_{l=-l_\mathrm{max}}^{l_\mathrm{max}}\left(\frac{cq^l}{\Omega_q^{\sigma}}\right)^2\Bigg\{ \ln{\left(\frac{\omega_\mathrm{c}}{c|q^{l}|}\right)} \nonumber \\
        &\, + \frac{1}{2}\left[ 1 + \mathrm{sgn}\left\{\eta^{\sigma}\right\} \left(\frac{\Omega_q^{\sigma}}{cq^l}\right)^2\right] \nonumber \\
        &\, \times \mathrm{log}^{\phi}\left(\frac{(cq^{l})^2 - {\Omega_q^{\sigma}}^2}{\omega_\mathrm{c}^2 - {\Omega_q^{\sigma}}^2}\right) \Bigg\}\Theta\left(\omega_\mathrm{c} - c|q^l|\right),
\end{align}
where we used the short form $q^l= q - 2\pi l/d$.  The use of a simple root-finding algorithm on Eq.~\eqref{eq:Eigenfreq_exact_developed} finally leads to the complex eigenfrequencies $\Omega_q^{\sigma}$.

Importantly, the logarithms $\mathrm{log}^{\phi}(z)$ present in the third line of the above equation are complex logarithms, and hence a choice $\phi$ of branch cut must be made. 
We recall that one can define any complex logarithm as
$    \mathrm{log}^{\phi}(z) = \ln(|z|) + \mathrm{i}\,\mathrm{arg}(z\mathrm{e}^{-\mathrm{i}\phi}) + \mathrm{i}\phi$,
with $\phi = \phi(|z|)$ being any real function of the modulus of $z$, $\ln(x)$ the natural logarithm and $\mathrm{Arg}(z)$ the principal value of the argument.
In our case, the use of the usual principal value logarithm, corresponding to the choice $\phi=0$, hides one of the roots behind the branch cut. 
To ensure that our algorithm can readily find the complex roots, and that they have a physical meaning, i.e., a positive (negative) real (imaginary) part [cf.\ Eq.~\eqref{eq:ansatz}], we have to modify the branch cut and we choose $\phi=-\pi/2$, leading to a logarithm which is continuous on the open set $\mathbb{C}\setminus \{\mathrm{i}\mathbb{R^+}\}$. 
Within such a choice, the root is no longer hidden and the algorithm can readily converge.

\subsection{Second-order perturbation theory}
\label{sec:Light Matter Interaction Perturbative}
The approach developed in the previous subsection allows us to take into account the retardation effects in an exact manner, diagonalizing the open quantum system Hamiltonian \eqref{eq:Hamiltonian} to find the complex eigenfrequencies of the system. 
However, it requires the numerical solution of the transcendental dispersion equation \eqref{eq:Eigenfreq_exact_developed}, in a similar way to the resolution using the classical approach (see Appendix~\ref{sec:AppendixClassicalModel}).
An advantage of our quantum approach is that we can use a perturbative treatment of the light--matter interaction, considering the interaction of the photonic continuum as a weak perturbation to the plasmonic system, to find adequate analytical approximations for the eigenfrequencies and the radiative decay rates of the system.

In this subsection, we rely on the work presented in Ref.~\cite{Downing2017}, with the substantial improvement of taking into account umklapp processes, which enables us to study both near- and far-field regimes.

\subsubsection{Radiative frequency shift}
We treat the coupling Hamiltonian \eqref{eq:H_plph} up to second order in perturbation theory.
For this purpose, we begin by diagonalizing the plasmonic Hamiltonian \eqref{eq:H_pl} by means of a Bogoliubov transformation. As detailed in Ref.~\cite{Downing2017}, one can rewrite the latter as $H^{\sigma}_{\mathrm{pl}}=\sum_q \hbar\omega_q^{\sigma}{B_q^{\sigma}}^{\dagger}B_q^{\sigma}$, where the quasistatic dispersion $\omega_q^{\sigma}$ is given in Eq.~\eqref{eq:quasistatic}, and the bosonic Bogoliubov ladder operator $B_q^{\sigma}$ (${B_q^{\sigma}}^{\dagger}$) acts on an eigenstate $\ket{n_q^{\sigma}}$ of the Hamiltonian \eqref{eq:fourier plasmonic Hamiltonian} representing $n_q^{\sigma}$ quanta occupying the collective plasmon
mode with polarization $\sigma$, wave vector $q$, and eigenfrequency $\omega_q^{\sigma}$, as $B_q^{\sigma}\ket{n_q^{\sigma}} = \sqrt{n_q^{\sigma}}\ket{n_q^{\sigma} - 1}$ (${B_q^{\sigma}}^{\dagger}\ket{n_q^{\sigma}} = \sqrt{n_q^{\sigma}+1}\ket{n_q^{\sigma} + 1}$). Explicitly, one has
\begin{align}
B_q^\sigma=\frac{\omega_q^\sigma+\omega_0}{2\sqrt{\omega_0\omega_q^\sigma}}\, b_q^\sigma
+\frac{\omega_q^\sigma-\omega_0}{2\sqrt{\omega_0\omega_q^\sigma}}\, {b_{-q}^\sigma}^\dagger.
\end{align}

For a given polarization $\sigma$ and mode $q$, the perturbed plasmonic energy levels are $E_{n_q^{\sigma}} = E_{n_q^{\sigma}}^{(0)} + E_{n_q^{\sigma}}^{(1)} + E_{n_q^{\sigma}}^{(2)}$. 
The renormalized plasmonic eigenfrequency within perturbation theory (pt) $\Omega_q^{\sigma,\mathrm{pt}}$ is defined as the difference between successive plasmonic energy levels, $\Omega_q^{\sigma,\mathrm{pt}} = (E_{n_q^{\sigma}+1}  - E_{n_q^{\sigma}})/\hbar $, and we write it as
\begin{equation}
\label{eq:Omega perturbative}
    \Omega_q^{\sigma,\mathrm{pt}} = \omega_q^{\sigma} + \delta_q^{\sigma}.
\end{equation}
Since the unperturbed energy is $E_{n_q^{\sigma}}^{(0)} = n_q^{\sigma}\hbar\omega_q^{\sigma}$ and the first-order contribution $E_{n_q^{\sigma}}^{(1)}=0$, the radiative frequency correction reads $\delta_q^{\sigma} = (E^{(2)}_{n_q^{\sigma}+1}  - E^{(2)}_{n_q^{\sigma}})/\hbar$ and can be written as
\begin{equation}
    \delta_q^{\sigma} = 2\omega_0^2\omega_q^{\sigma}\frac{\pi a^3}{dL^2}\sum_l\sum_{\boldsymbol{\kappa},\hat{\lambda}^l_{\boldsymbol{\kappa}q}}\frac{( \hat{\sigma}\cdot \hat{\lambda}^l_{\boldsymbol{\kappa}q} )^2\;\Theta\left(\omega_\mathrm{c} - \nu^l_{\boldsymbol{\kappa}q}\right)}{{\omega_q^{\sigma}}^2 - {\nu^l_{\boldsymbol{\kappa}q}}^2},
\end{equation}
where the summation over $\boldsymbol{\kappa}$ excludes singular terms. We note that the only correction to the plasmonic energy levels which contributes is the second-order one, which corresponds to the emission and reabsorption of virtual photons by the plasmonic eigenstate $\ket{n_q^{\sigma}}$.

Similarly as in the previous subsection, one can carry out the summation over photon polarization using the relation \eqref{eq:polarization sum} and taking the continuum limit \eqref{eq:continuum limit} to write the summation over $\boldsymbol{\kappa}$ as a principal value integral using polar coordinates. 
We then obtain
\begin{align}
        \label{eq:radiative shift}
       \delta_q^{\sigma} =&\,       \eta^{\sigma}\frac{\omega_0^2{\omega_q^{\sigma}}}{2}\frac{a^3}{dc^2}\sum_l\Theta\left(\omega_\mathrm{c} - c|q^l|\right)\left(\frac{cq^l}{\omega_q^{\sigma}}\right)^2\Bigg\{ \ln{\left(\frac{\omega_\mathrm{c}}{c|q^{l}|}\right)} \nonumber \\
        &\, + \frac{1}{2}\left[ 1 + \mathrm{sgn}\left\{\eta^{\sigma}\right\} \left(\frac{\omega_q^{\sigma}}{cq^l}\right)^2\right] \mathrm{ln}\left(\left|\frac{(cq^{l})^2 - {\omega_q^{\sigma}}^2}{\omega_\mathrm{c}^2 - {\omega_q^{\sigma}}^2}\right|\right) \Bigg\}.
\end{align}
Note that the above radiative frequency shift, aside from a factor of $2\omega_q^{\sigma}$, has the same expression as the right-hand side of  Eq.~\eqref{eq:Eigenfreq_exact_developed}, except that the complex polaritonic eigenfrequency $\Omega_q^{\sigma}$ has been replaced by the quasistatic plasmonic frequency $\omega_q^{\sigma}$, and hence the complex logarithms have became natural logarithms.
We further note that due to the plasmon--photon umklapp processes, and unlike what has been done in previous works \cite{BrandstetterKunc,Downing2017}, the $2\pi/d$-periodicity of the eigenfrequencies $\Omega_q^{\sigma,\mathrm{pt}}$ is conserved.

\subsubsection{Radiative decay rate}

Treating the light--matter coupling Hamiltonian \eqref{eq:H_plph} as a weak perturbation, one can use Fermi's golden rule to obtain the radiative decay rates $\gamma_q^{\sigma,\mathrm{pt}}$ of the collective plasmonic excitation into the surrounding photonic environment. Using the expression \eqref{eq:fourier coupling Hamiltonian} of the coupling Hamiltonian, we find that these decay rates, which also represent the radiative contribution to the inverse lifetime of a given mode $q$, read
\begin{align}
    \gamma_q^{\sigma,\mathrm{pt}} =&\, 2\pi^2\omega_0^2\omega_q^{\sigma}\frac{a^3}{dL^2}\sum_l\sum_{\boldsymbol{\kappa},\hat{\lambda}^l_{\boldsymbol{\kappa}q}}\frac{( \hat{\sigma}\cdot\hat{\lambda}^l_{\boldsymbol{\kappa}q})^2}{\nu^l_{\boldsymbol{\kappa}q}} \delta\left(\omega_q^{\sigma} - \nu^l_{\boldsymbol{\kappa}q} \right) \nonumber \\
    &\, \times\Theta\left( \omega_\mathrm{c} - \nu^l_{\boldsymbol{\kappa}q} \right).
\end{align}
After taking the continuum limit \eqref{eq:continuum limit} and completing the integration, the above equation can be written as
\begin{align}
\label{eq:decay rate fermi's golden rule}
    \gamma_q^{\sigma,\mathrm{pt}} =&\, \frac{\pi\eta^{\sigma}}{2}\frac{\omega_0^2}{\omega_q^{\sigma}}\frac{a^3}{d}\sum_l \left( q^l \right)^2 \left[ 1 + \mathrm{sgn}\{\eta^{\sigma}\}\left( \frac{\omega_q^{\sigma}}{cq^l} \right)^2 \right] \nonumber \\
    &\, \times \Theta\left(\omega_q^{\sigma} - c|q^l|\right).
\end{align}

\section{Results}\label{sec:Results}

We now present the results obtained through our numerical quantum approach presented in Sec.~\ref{sec:Light Matter Interaction Hopfield}, which we compare to the analytical perturbative results presented in Sec.~\ref{sec:Light Matter Interaction Perturbative} and to the classical model widely used in the literature and outlined in Appendix~\ref{sec:AppendixClassicalModel}.
Specifically, the classical model which we employ uses the nanoparticle polarizability of Eq.~\eqref{eq:Polarizability}
that is obtained from the first (electric dipole) Mie coefficient \cite{Mie,Bohren2004,Capolino2009}, see Eq.~\eqref{eq:Mie_coeff}.

We begin this section by displaying our results for the polaritonic band structures and the radiative decay rates (Sec.~\ref{sec:Dispersion and decay rate}) and then discuss the corresponding group velocities (Sec.~\ref{sec:Group velocities}). We then quantitatively demonstrate the hybridization of the light and matter degrees of freedom through a detailed discussion of the Hopfield coefficients (Sec.~\ref{sec:Hopfield}), and, finally, we discuss the limit of infinitely spaced nanoparticles, i.e., of single nanoparticles (Sec.~\ref{sec:Single NP Limit}).

\subsection{Polaritonic band structure and radiative decay rate}\label{sec:Dispersion and decay rate}

In this subsection, we present the results obtained from the numerical resolution of Eq.~\eqref{eq:Eigenfreq_exact_developed} for the band structure $\mathrm{Re}\{\Omega^{\sigma}_q\}$ and the radiative decay rate $\gamma_q^{\sigma} = -2\mathrm{Im}\{\Omega^{\sigma}_q\}$, for the longitudinal ($\sigma=z$) and transverse ($\sigma=x,y$) polarizations. The factor $-2$ in the above expression of the decay rate is such that $\gamma_q^{\sigma}$ corresponds to the radiative contribution to the inverse lifetime of a given mode of the system [see Eq.~\eqref{eq:ansatz}].\footnote{Explicitly, with the decomposition of $\Omega^{\sigma}_q$ into real and imaginary parts mentioned above, the polaritonic annihilation operator \eqref{eq:ansatz} takes the form $\mu_q^\sigma(t)=\mu_q^\sigma\,\mathrm{e}^{-\mathrm{i}\mathrm{Re}\{\Omega^{\sigma}_q\}t}\,\mathrm{e}^{-\gamma_q^\sigma t/2}$, such that the decay rate of the collective excitation corresponds to $\gamma_q^\sigma$.}
In Figs.~\ref{fig:dispersion}--\ref{fig:dispersion_23a} below, we compare these results with the perturbative ones, given by Eqs.~\eqref{eq:Omega perturbative} and \eqref{eq:decay rate fermi's golden rule}, and to the results from the classical model obtained through the numerical resolution of Eq.~\eqref{eq:Classical dispersion}.

We choose a value of $k_0a = 0.3$ which corresponds to, e.g., nanoparticles with a Mie frequency $\omega_0 = \unit[3]{\mathrm{eV}/\hbar}$ and a radius $a=\unit[20]{nm}$. We display the eigenfrequencies and decay rates in units of the bare frequency $\omega_0$ and as a function of the reduced longitudinal wave number $qd$. In this way, we consider $\omega_0$ and hence the plasma frequency $\omega_{\mathrm{p}}$ as (essentially material-dependent) parameters of our theory.
Only the positive half of the first Brillouin zone is shown, since all of the plotted quantities are symmetric around $q=0$ and $2\pi/d$-periodic.

\subsubsection{Near-field coupled nanoparticles}

\begin{figure*}[t]
    \includegraphics[width=\linewidth]{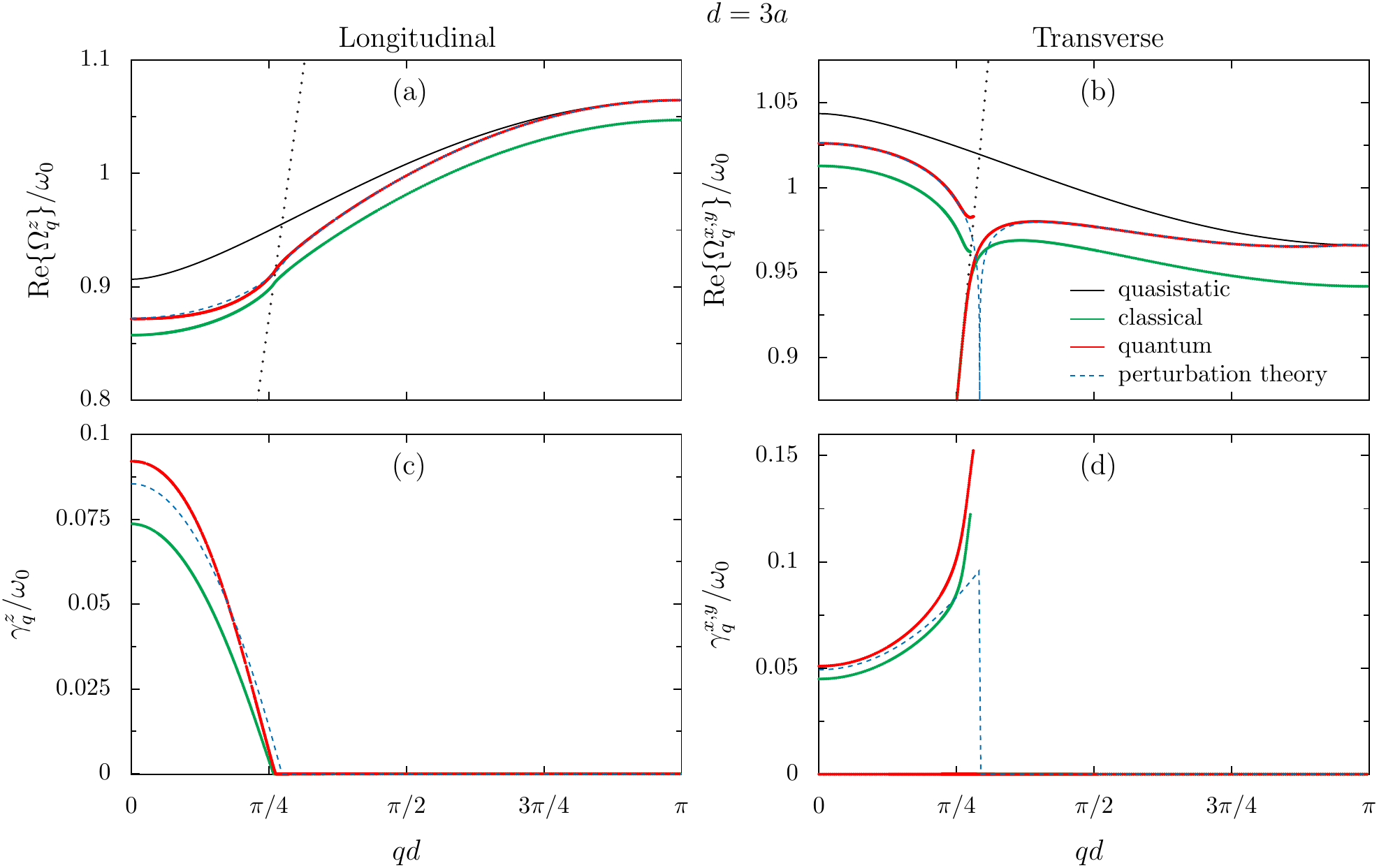}
    \caption{(a), (b) Band structure and (c), (d) radiative decay rates in units of the bare frequency $\omega_0$ and as a function of the reduced wave number $qd$ in half of the Brillouin zone. The left (right) panels represent longitudinal (transverse) polarizations.
    The red curves correspond to the results obtained within our fully retarded quantum formalism while the green ones are the results obtained within the classical model presented in Appendix~\ref{sec:AppendixClassicalModel}. The blue dashed lines represent the second-order perturbation theory and Fermi's golden rule analytical results. In the upper panels, we also display the quasistatic band structure \eqref{eq:quasistatic} by solid black lines and the first ($l=0$) light line, $c|q|$, as a dotted black line. The center-to-center distance is $d=3a$ and $k_0 a=0.3$.
    }
    \label{fig:dispersion}
\end{figure*}

We first discuss in Fig.~\ref{fig:dispersion} the case of a center-to-center distance $d=3a$. For such a separation, the LSPs supported by the nanoparticles are near-field coupled.

Compared to the quasistatic dispersion from Eq.~\eqref{eq:quasistatic} [black solid lines in Figs.~\ref{fig:dispersion}(a) and \ref{fig:dispersion}(b)], the fully retarded one (red solid lines) experiences a redshift, especially in the area inside the first light cone, i.e., for frequencies larger than the first ($l = 0$) light line $c|q|$ (depicted by a black dotted line). 
In this area of the Brillouin zone, the light--matter interaction is significant and the plasmons are radiating. 
Thus, a finite radiative linewidth appears [see Figs.~\ref{fig:dispersion}(c) and \ref{fig:dispersion}(d)], which vanishes at the wave number where the dispersion relation intersects the light line. 
We note that at this precise point, the dispersion curve for the longitudinal polarization [see Fig.~\ref{fig:dispersion}(a)] presents a discontinuity, i.e., no roots are found. Taking into account Ohmic losses, however, leads to a definite root at the intersection \cite{Compaijen,Jacak2014}, where the slope of the curve changes drastically to be equal to the slope of the light line, namely, $c$.
For larger values of $qd$, the plasmonic excitation is interacting with photonic modes of higher energy and the normal modes correspond to guided ones, which are immune to radiation damping.

From the results displayed in  Figs.~\ref{fig:dispersion}(a) and \ref{fig:dispersion}(c), we see that for the longitudinal polarization there are almost no differences between our fully retarded eigenfrequencies obtained from Eq.~\eqref{eq:Eigenfreq_exact_developed} and the analytical second-order perturbation result \eqref{eq:Omega perturbative} (dashed blue lines). 
Likewise, we only have a slight increase of the decay rate $\gamma^z_q$ in comparison with the Fermi golden rule result [Fig.~\ref{fig:dispersion}(c)].
The longitudinally polarized plasmonic chain is thus very well described by our analytical theory.
This is due to the fact that in the Coulomb gauge, only transverse photons are present and hence they interact weakly with the longitudinally polarized plasmons. 

For the transverse polarizations [Figs.~\ref{fig:dispersion}(b) and \ref{fig:dispersion}(d)], the full treatment of the light--matter interaction implies the appearance of two distinct bands in the dispersion, anticrossing at the light line. This is a typical avoided crossing dispersion, which is a signature of the presence of polaritonic excitations \cite{Hopfield}. Here and in the following, we denote as the upper band the one which is above the light line anticrossing the dispersion (here $c|q|$), and the lower band the one below. We see in Figs.~\ref{fig:dispersion}(b) and \ref{fig:dispersion}(d) that for $d=3a$, the upper band represents modes with radiating polaritons inside the light cone [i.e., with a finite decay rate, cf.\ the upper red line in Fig.~\ref{fig:dispersion}(d)] and stops at the anticrossing, while the lower band is free of radiation damping 
in the entire Brillouin zone
[red line corresponding to $\gamma_q^{x,y}=0$ in Fig.~\ref{fig:dispersion}(d)] and is therefore a real solution of Eq.~\eqref{eq:Eigenfreq_exact_developed}.
This polaritonic behavior is not described by the result of perturbation theory, which is nevertheless very close to the fully-retarded result aside from the region around the anticrossing with the light line, which appears at a wave number $q \simeq k_0$. 

By comparing the above quantum results to the classical ones (green solid lines in Fig.~\ref{fig:dispersion}), one notices that the redshift in the dispersion 
[Figs.~\ref{fig:dispersion}(a) and \ref{fig:dispersion}(b)] is stronger ($1$--$\unit[3]{\%}$) in the case of the classical computation. A significant difference between the two models is present for the radiative decay rate only, where the classical model predicts a rate up to $\unit[25]{\%}$ lower than the quantum model. However, if the relative difference is noticeable, the absolute one is very small with a maximum deviation of $0.03\,\omega_0$.

\begin{figure*}[tb]
    \includegraphics[width=\linewidth]{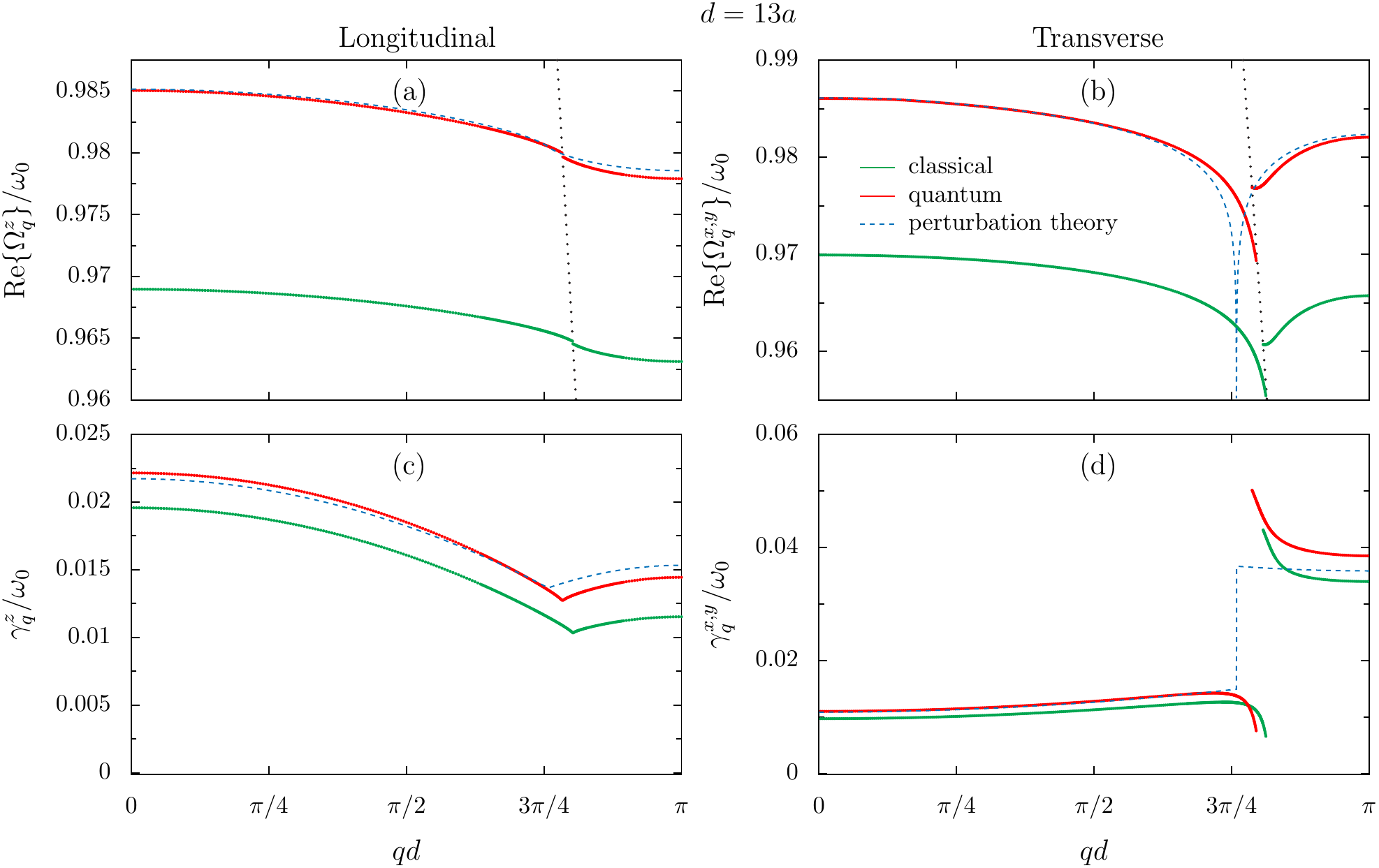}
    \caption{Same quantities as shown in Fig.~\ref{fig:dispersion}, except for the black dotted line in (a) and (b), which is now the light line corresponding to the photonic band index $l=1$, i.e., $c|q-2\pi/d|$. In the figure, $k_0a = 0.3$ and $d = 13a$.}
    \label{fig:dispersion_13a}
\end{figure*}

As experienced by other authors \cite{Koenderink,Fung2007}, we do not find data points of the retarded band structure for the upper band after the anticrossing with the light line, in contrast with the usual picture of the avoided crossing where it extends upward. It has been argued that the large losses in this region prevent one from finding well-defined roots \cite{Koenderink}.
We also notice that in the transverse case the real roots, corresponding to the lower band, are difficult to find for values of $qd \ll 1$. 
The problem is even more stringent using the classical model, where no solutions can be found \cite{Koenderink,Compaijen}. 

Finally, we remark that for values of $qd\geqslant3$, the fully retarded and quasistatic dispersion relations become indistinguishable. This is due to the ultraviolet cutoff $\omega_\mathrm{c}$ we chose, which limits our theory to wave numbers $|q-2\pi l/d|\leqslant k_\mathrm{c}=1/a$ and is consistent with the fact that at the edge of the Brillouin zone, the plasmonic excitations are interacting with photonic modes with energy three times larger, which implies weak coupling.
Within this choice of cutoff, only the first light line $l=0$ is of sufficiently low energy to interact with the plasmonic excitations when the center-to-center distance is set to a value of $d$ such that $d\leqslant\pi a$. This explains why the perturbative study of the case $d=3a$ can be carried out without taking into account the umklapp processes \cite{Downing2017}.

\subsubsection{Far-field coupled nanoparticles}

Considering now a larger center-to-center distance $d=13a$, far-field effects take place and umklapp processes become of great importance.
We show in Fig.~\ref{fig:dispersion_13a} the same quantities as previously, i.e., the dispersion relation for the longitudinal [Fig.~\ref{fig:dispersion_13a}(a)] and transverse [Fig.~\ref{fig:dispersion_13a}(b)] polarizations, and the corresponding radiative linewidths in Fig.~\ref{fig:dispersion_13a}(c) and Fig.~\ref{fig:dispersion_13a}(d), respectively.
We see that the same conclusions as for the case $d=3a$ of Fig.~\ref{fig:dispersion} can be drawn from the comparison between the different models, namely, that the perturbation theory reproduces, at least qualitatively in the transverse direction, the exact quantum result for the polaritonic band structure, which itself is close to the results obtained through a purely classical model. As before, however, the perturbation theory misses the avoided crossing feature of the transverse polarization and hence is not reliable around the anticrossing. We also note that since the dipolar coupling strength $\Omega$ scales as $(a/d)^3$ [cf.\ Eq.~\eqref{eq:Omega}], the bandwidth is significantly reduced when we increase the spacing between the nanoparticles. This leads the quasistatic dispersion \eqref{eq:quasistatic} to being reduced to almost $\omega_0$ in the entire Brillouin zone, so we no longer show it in Figs.~\ref{fig:dispersion_13a}(a) and \ref{fig:dispersion_13a}(b).

What is also immediately remarkable by comparing Figs.~\ref{fig:dispersion} and \ref{fig:dispersion_13a} is the flipping of the dispersion curves, the slope of the longitudinal dispersion curve being now negative, as well as the upper band in the transverse case.
This is due to the fact that here, the longitudinal (transverse) dispersion intersects (anticrosses) the light line corresponding to the photonic band index $l=1$, i.e., the curve  $c|q-2\pi/d|$, which is also decreasing in the positive half of the first Brillouin zone, as $qd$ increases. 
The slope of the closest light line hence dictates the one of the longitudinal dispersion, and as we increase the center-to-center distance $d$, the curve continuously evolves from an increasing
[Fig.~\ref{fig:dispersion}(a)] to a decreasing 
[Fig.~\ref{fig:dispersion_13a}(a)] function, and similarly for the transverse polarization [Figs.~\ref{fig:dispersion}(b) and \ref{fig:dispersion_13a}(b)].
We note that we still observe a discontinuity in the longitudinal dispersion at the intersection with the light line, and that in the transverse case, no roots can be found along a light line $l\neq0$, neither upward nor downward, for both the quantum or classical approaches.

\begin{figure*}[tb]
    \includegraphics[width=\linewidth]{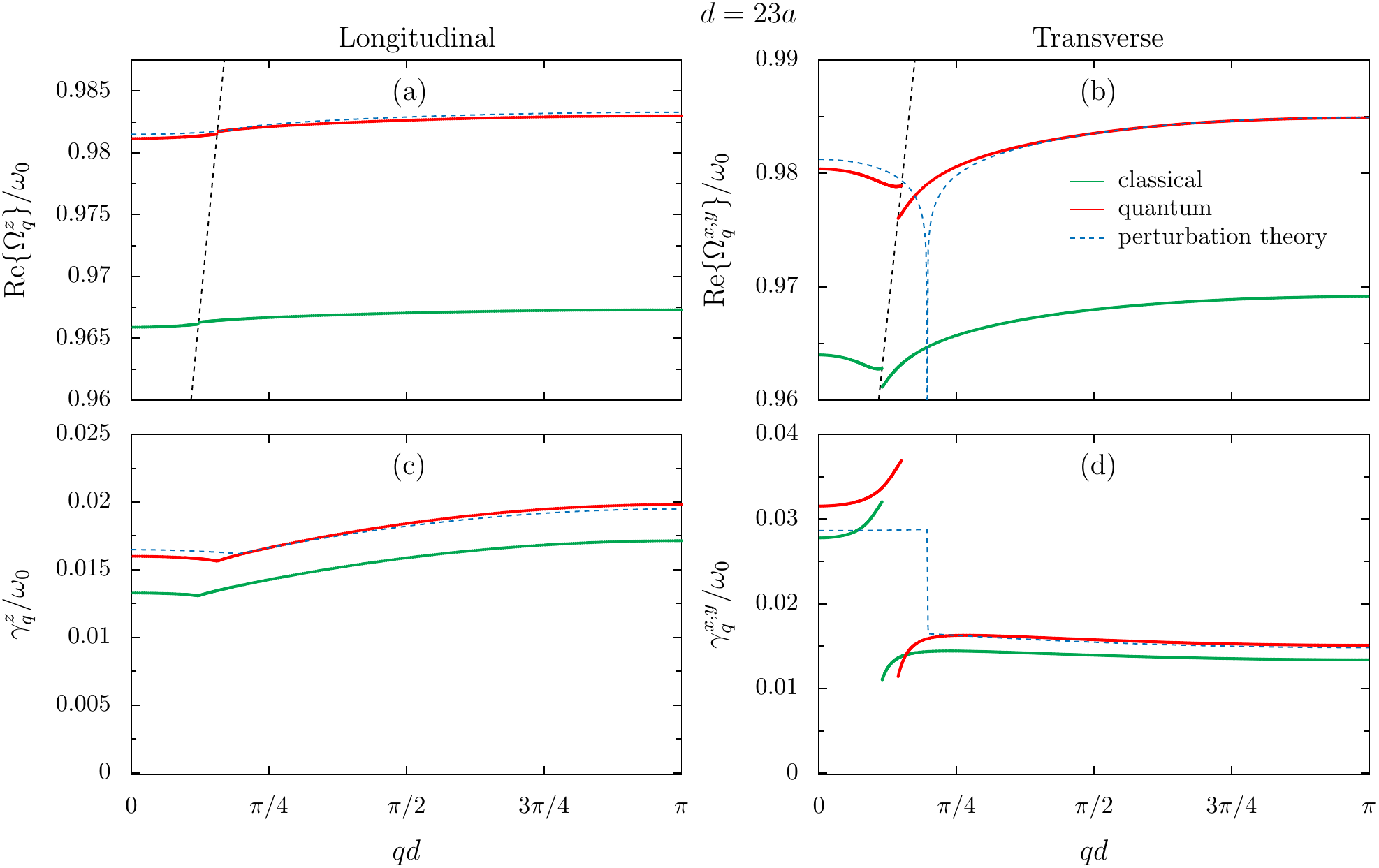}
    \caption{Same quantities as shown in Figs.~\ref{fig:dispersion} and \ref{fig:dispersion_13a}, except for the black dotted line in panels (a) and (b), which is now the light line corresponding to the photonic band index $l=-1$, i.e., $c|q+2\pi/d|$. In the figure, $k_0a = 0.3$ and $d = 23a$.}
    \label{fig:dispersion_23a}
\end{figure*}

Since all the dispersion curve is included into the first light cone, i.e., is above the first light line $c|q|$, all the modes are now radiating, as can be seen in Figs.~\ref{fig:dispersion_13a}(c) and \ref{fig:dispersion_13a}(d). This is in stark contrast with the case $d=3a$ where guided modes immune to radiation damping are present [Figs.~\ref{fig:dispersion}(c) and \ref{fig:dispersion}(d)]. 
The latter guided modes disappear completely, starting from a center-to-center distance of around $d\simeq11a$, for longitudinal (transverse) polarization and in both classical and quantum models, when the dispersion curve intersects (anticrosses) the first light line $c|q|$ at the edge of the Brillouin zone.
Interestingly, this means that for a specific distance $d\simeq11a$, the dispersion curve will intersect both the first ($l=0$) and second ($l=1$) light lines. For this precise interparticle distance, one observes a double avoided-crossing scheme in the transverse case, leading to three distinct eigenfrequencies for a given mode $q$.

The second light line $l=1$ also forms a light cone, containing the modes $q$ whose frequencies are larger than $c|q-2\pi/d|$, that is to say, the modes on the right of the black dotted line in Figs.~\ref{fig:dispersion_13a}(a) and \ref{fig:dispersion_13a}(b). 
For the radiative linewidth in the longitudinal case [Fig.~\ref{fig:dispersion_13a}(c)], one can observe a cusp at the point where the dispersion intersects the second light line, followed by a slight increase of the linewidth. More interestingly, in the transverse case [Fig.~\ref{fig:dispersion_13a}(d)], we see that the effect of the second light cone is to enhance the radiative linewidth. The two bands hence radiate very differently, the one inside the second light cone having a radiative linewidth almost four times larger than the one outside. We note that for the modes around the anticrossing, the perturbation theory misses the decrease (increase) of the radiative linewidth of the band outside (inside) the second light cone, and approximates the behavior to an almost step-like function.

Finally, we consider the case of $d=23a$ in Fig.~\ref{fig:dispersion_23a}. 
Now the intersecting light line is the one with band index $l=-1$, i.e., the curve $c|q+2\pi/d|$ [see the black dotted lines in Figs.~\ref{fig:dispersion_23a}(a) and \ref{fig:dispersion_23a}(b)].
Therefore, the dispersion curves are once again flipped and we again see a different radiating behavior [Figs.~\ref{fig:dispersion_23a}(c) and \ref{fig:dispersion_23a}(d)], depending on whether the polaritonic band is inside or outside the light cone formed by the third photonic band $l=-1$.

We also notice from Fig.~\ref{fig:dispersion_23a} that both the dispersion relations and radiative decay rates tend to become flatter. 
This is consistent with the fact that when the center-to-center distance tends to infinity, the nanoparticles become isolated from each other and there is no more collective effects along the chain. 
Therefore, all the eigenfrequencies must become degenerate to the value corresponding to the excitation of a single nanoparticle.
We will come back in detail to this limiting case in Sec.~\ref{sec:Single NP Limit}.

\subsection{Group velocity of the polaritonic modes} \label{sec:Group velocities}
\begin{figure}[tb]
    \includegraphics[width=\columnwidth]{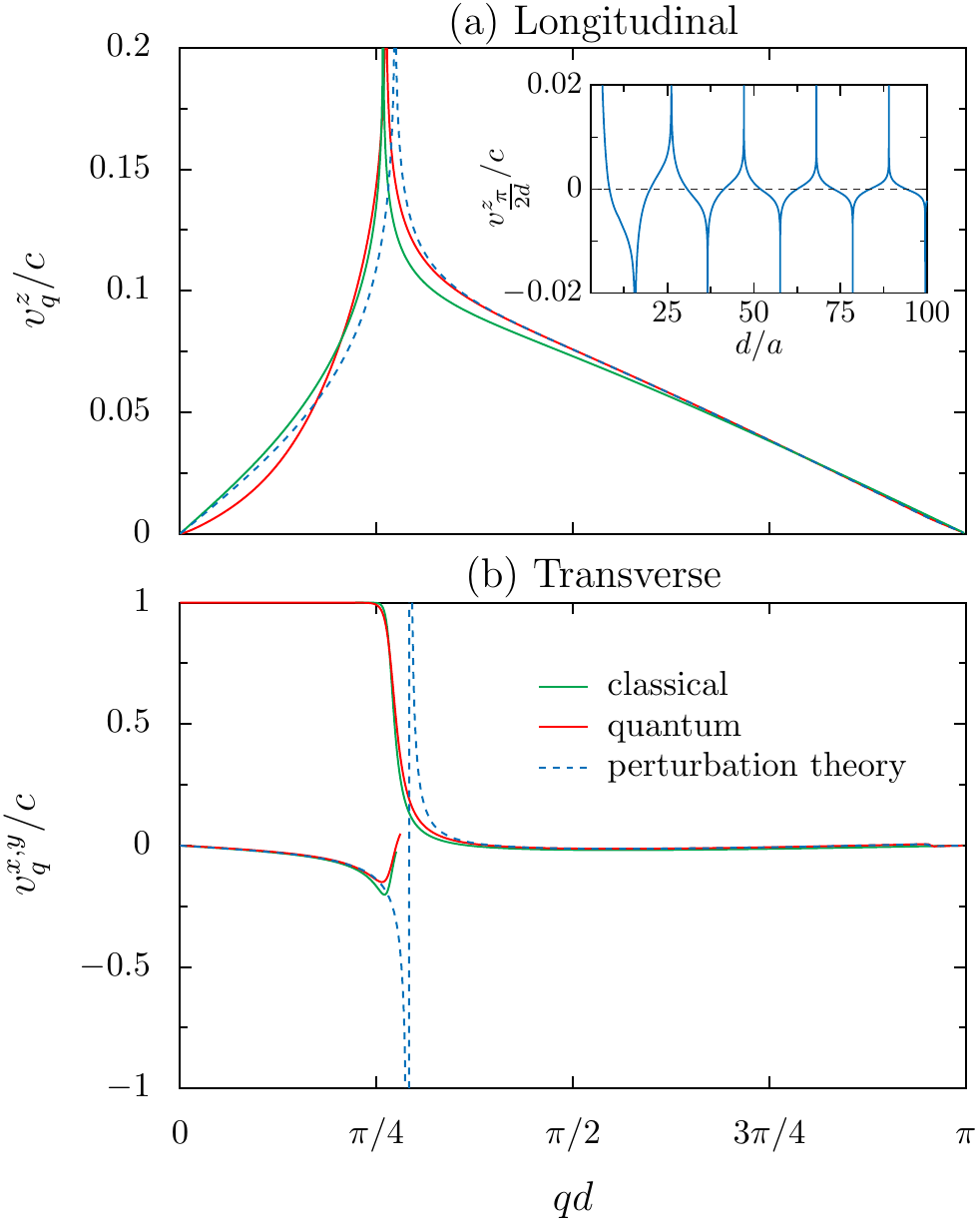}
    \caption{Group velocities $v^{\sigma}_q$ in units of the speed of light in vacuum $c$, for the (a) longitudinal and (b) transverse polarizations as a function of the reduced wave number $qd$ in half of the first Brillouin zone. The parameters used in the figure are the same as in Fig.~\ref{fig:dispersion}. The inset shows the longitudinal group velocity in units of $c$ corresponding to the reduced wave number $qd=\pi/2$ as a function of the ratio $d/a$, computed using the analytical perturbative results.}
\label{fig:Velocities}
\end{figure}

To insight into the propagation of the polaritonic modes, we define their group velocity as the derivative of the dispersion, $v^{\sigma}_q = \partial \mathrm{Re}\{\Omega_q^{\sigma}\} / \partial q$. Figure~\ref{fig:Velocities} shows the results obtained using a center-to-center distance $d=3a$, i.e., by taking the derivatives of the curves displayed in 
Figs.~\ref{fig:dispersion}(a) and \ref{fig:dispersion}(b). 
We first note that as for the dispersion relations, the group velocities are in very good agreement between the classical and quantum models, for both polarizations.

Unlike what could be expected from its dispersion, the longitudinal case admits quite large group velocities, with a fair amount of modes propagating at almost $0.1c$ or faster, providing a notable window for propagation just after the intersection with the light line, where guided modes present no radiative damping and a rather large group velocity.

At the precise intersection with the light line, as discussed before, no definite value is found in our model free of Ohmic losses, and we observe huge variations of the derivative around the intersection [see 
Fig.~\ref{fig:Velocities}(a)], implying difficulties to obtain a good numerical accuracy for values of $v_q^{z}>0.2c$. Taking into account Ohmic losses, however, removes the discontinuity and leads to a group velocity of exactly $c$ at the intersection, as demonstrated in Refs.~\cite{Jacak2014,Compaijen} using two different approaches.
The analytical second-order computation only qualitatively accounts for this effect with the appearance of a logarithmic singularity in the group velocity at $\omega_q^{\sigma} = c|q - 2\pi l/d|$, as can be seen taking the derivative of Eq.~\eqref{eq:radiative shift}.
This unphysical infinite group velocity at the crossing with the light line shows the limits of the analytical computation in the longitudinal case.

For the transversely polarized excitations [see Fig.~\ref{fig:Velocities}(b)], we see that the upper band in the dispersion, i.e., the radiating one, shows a rather large negative group velocity with a maximum of about $-0.15c$. The lower band, i.e., the guided modes, which follows the photonic dispersion $c|q|$ inside the light cone [see Fig.~\ref{fig:dispersion}(b)], propagates at the speed of light $c$ and then rapidly decreases around the intersection with the light line, where a very slow but notably slightly negative group velocity is found.
The perturbative treatment, once again, reaches its limitations and presents an unphysical singularity at the crossing.

When we increase the center-to-center distance $d$, as observed in the previous subsection, the dispersion curves maintain a similar shape but get flatter, hence the group velocity globally decreases. 
However, the longitudinal modes will keep a large group velocity around the intersection with the light line, which has been found to be of exactly $c$ at the precise intersection when Ohmic losses are taken into account \cite{Jacak2014,Compaijen}.
Interestingly, as discussed previously, increasing the distance between the nanoparticles also leads to a continuous reversing of the dispersion curves, implying an inversion of the sign of the group velocities.

In the inset of Fig.~\ref{fig:Velocities}, we show the longitudinal group velocity computed from the perturbative results and associated to the mode with wave number $q=\pi/2d$, in units of the speed of light $c$ and as a function of the reduced center-to-center distance $d/a$.
Remarkably, we observe an almost periodic sign change of the group velocity. 
One can show that when we increase the ratio $d/a$, the pseudoperiod of the sign change tends very quickly to $\pi/k_0a$ in units of $d/a$, i.e., to $\sim 10.5$ within our choice of $k_0a=0.3$. 
In units of $d$, this amounts to an approximate period of $\lambda_0/2$, where $\lambda_0=2\pi/k_0$ is the resonance wavelength associated with the isolated single nanoparticle.
In the limit of $d\gg a$, the group velocity is hence almost periodic with a period $\lambda_0$.
This occurs for all modes $q$, except for $q=0$ and $q=\pm\pi/d$, respectively the center and edges of the Brillouin zone, for which the group velocity remains zero regardless of the interparticle distance $d$.
Interestingly, even if the first sign change appears at a different distance $d$ depending on the mode $q$, when $d\gg a$ the interparticle distance $r_n$ for which a sign change appears becomes approximately $r_n = n\lambda_0/2$, where $n$ is a positive integer.
More precisely, the roots of the longitudinal group velocity are given by $r_{q,n}(d) = n\lambda_0/2 + s_q(d)$, where $s_q(d) \ll \lambda_0/2$ is a rapidly decreasing shift that tends to $0$.
We also note that for the particular mode $q=\pi/2d$ shown in the inset, the absolute value of the group velocity is maximal (and could be equal to $\pm c$ in a model where Ohmic losses would be taken into account \cite{Jacak2014,Compaijen}) exactly for $r_n=(2n+1)\lambda_0/2 \mp \lambda_0/4$.

Importantly, the above-discussed unusual behavior of the group velocity allows for a high tunability, since all given modes $q$ except the ones at the center and edges of the Brillouin zone are associated with a wide range of different group velocities, including negative and vanishing ones, depending on the center-to-center distance chosen experimentally.
This phenomenon can be compared with the behavior of surface lattice resonances, which admit a strong narrowing of the plasmon resonance due to diffraction and interference effects when the distance between nanoparticle equals the resonance wavelength \cite{Auguie2008,Kravets2018}.

\subsection{Hybridization of light and matter degrees of freedom}\label{sec:Hopfield}

\begin{figure*}[tb]
    \includegraphics[width=.83\linewidth]{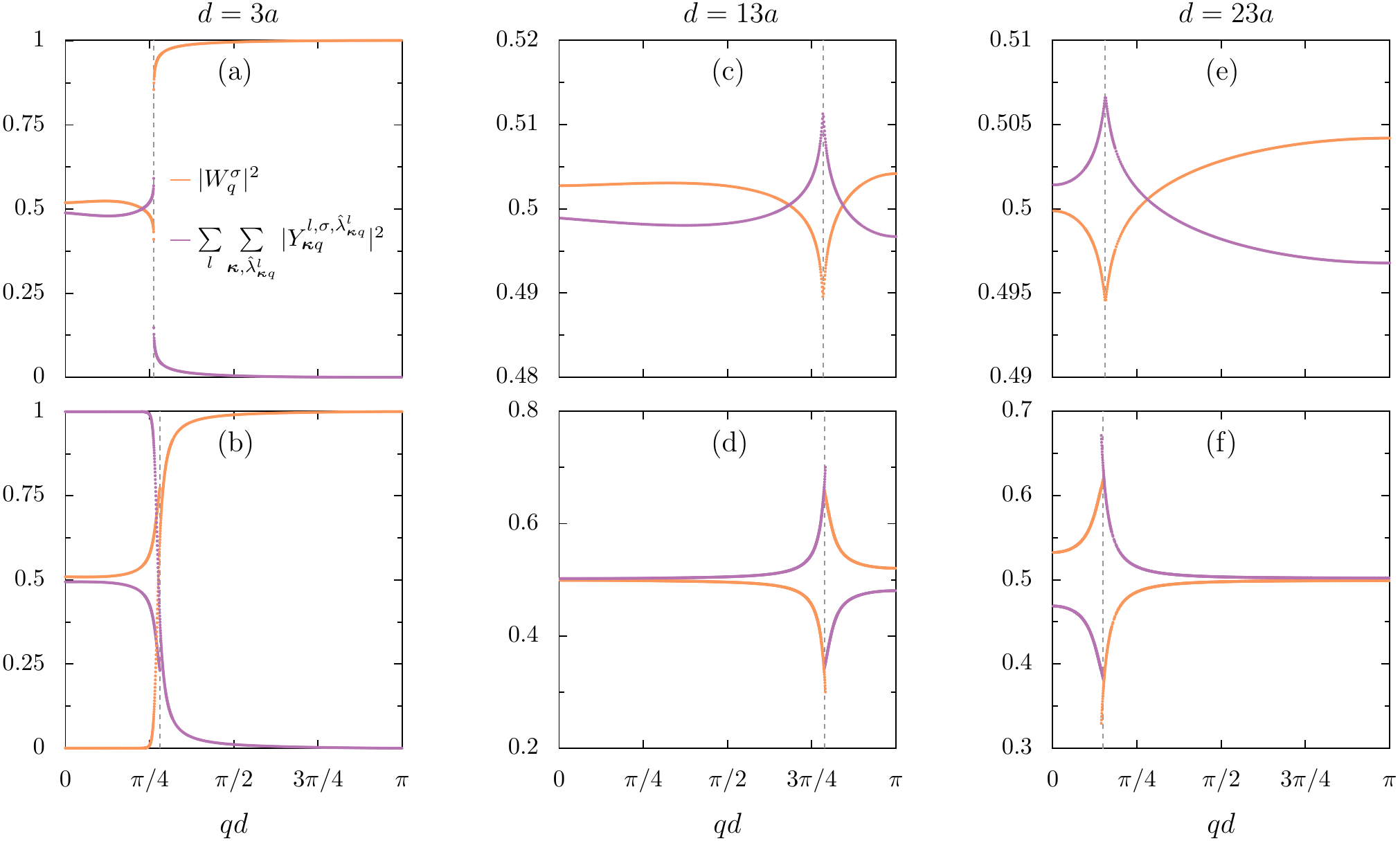}
    \caption{Modulus squared of the plasmonic Hopfield coefficients $W_{q}^{\sigma}$ (orange solid line) and sum of all the moduli squared of the photonic ones $Y_{\boldsymbol{\kappa}q}^{l,\sigma,\hat\lambda^l_{\boldsymbol{\kappa}q}}$ (purple solid line), as a function of the reduced wave number $qd$ in half of the Brillouin zone. The vertical dashed line represents the mode for which the polaritonic dispersion crosses a light line. The upper and lower panels show, respectively, the longitudinal and transverse polarizations, while the three columns are associated, respectively, to center-to-center distances $d=3a$, $d=13a$, and $d=23a$. In the figure, $k_0a= 0.3$.}
\label{fig:Hopfield}
\end{figure*}

As previously discussed, the plasmonic excitations hybridize with the photonic modes and form plasmon--polaritons, which are directly apparent through the avoided-crossing feature of the transverse dispersion relations.
Our quantum theory gives access to the share of plasmonic and photonic excitations within a given collective mode with wave number $q$.
This distribution can be found through the Hopfield coefficients introduced in Eq.~\eqref{eq:polaritonic operator}, where each coefficient represents the weighting of the plasmonic and photonic excitations.

Using the normalization condition \eqref{eq:normalization}, one can solve the system of equations \eqref{eq:system of equations}, respectively, for the moduli squared of the two plasmonic Hopfield coefficients and for the sum over all band indexes $l$, wave vectors $\boldsymbol{\kappa}$ and polarizations $\hat\lambda^l_{\boldsymbol{\kappa} q}$ of the moduli squared of the two photonic coefficients. 
Transforming the sums over $\boldsymbol{\kappa}$ into finite integrals, one obtains functions of the complex eigenfrequencies $\Omega_q^{\sigma}$. 
We give details on the computation and provide the analytical expressions of the Hopfield coefficients in Appendix~\ref{sec:AppendixHopfield}.

Using the eigenfrequencies found via Eq.~\eqref{eq:Eigenfreq_exact_developed}, we show in Fig.~\ref{fig:Hopfield} the results for the longitudinal (upper panels) and transverse (lower panels) polarizations, for center-to-center distances $d=3a$, $d=13a$, and $d=23a$.
We note that the two Hopfield coefficients $X_{q}^{\sigma}$ and $Z_{\boldsymbol{\kappa}q}^{l,\sigma,\hat\lambda^l_{\boldsymbol{\kappa}q}}$, associated with the counter-rotating terms in the polaritonic ladder operator \eqref{eq:polaritonic operator} are not plotted in the figure, since their largest contributions are less than $0.006$. 
This suggests that we could have employed the rotating wave approximation (RWA) in our derivations. 
However, since the RWA does not lead to any computational simplification, we chose to use the full Hamiltonian \eqref{eq:fourier polaritonic Hamiltonian}.

Remarkably, we notice that for the longitudinal polarization [Figs.~\ref{fig:Hopfield}(a), \ref{fig:Hopfield}(c), and \ref{fig:Hopfield}(e)], and hence without having any avoided-crossing dispersion, the normal modes inside a light cone are an equitable mix of light and matter degrees of freedom, proving that we do indeed have plasmon--polaritons, even in this case.
In the near-field case [Fig.~\ref{fig:Hopfield}(a), $d=3a$] at the end of the light cone, the excitation becomes almost purely plasmonic, which is consistent with the suppression of radiative losses. 
For center-to-center distances such that the dispersion curve intersects a photonic band of index $l\neq0$ [Figs.~\ref{fig:Hopfield}(c) and \ref{fig:Hopfield}(e)], the entire Brillouin zone is located within the light cone $l=0$ and all modes are radiating. 
Thus in this case all modes are polaritonic.
Approaching the mode for which the dispersion curve crosses the light line, represented as a vertical dashed line in the figure, the excitation becomes mostly photonic. 
At the exact crossing, the Hopfield coefficients are not rigorously defined due to the discontinuity and avoided crossing in the dispersions, but one can argue that at this precise point, the excitation tends to be purely photonic, consistent with the fact that the associated group velocity tends to $\pm c$.

For the transverse polarization, let us first discuss the near-field case, Fig.~\ref{fig:Hopfield}(b). The upper plasmon--polariton band, associated with a nonzero radiative damping, presents a strong part of photon excitation which, quite surprisingly, decreases when approaching the intersection with the light line.
We note that the polaritonic nature is marked for all modes inside the light cone, and not only those close to the intersection. 
This can be understood by the fact that plasmons interact with all photonic modes $\mathbf{k}$ and not only with the light line, which corresponds to $\mathbf{k}=(0,0,q)$.
We further notice that the lower band of the transverse dispersion is almost exclusively composed of photons inside the light cone, which is in adequacy with the fact that it is not subject to radiative losses and has a dispersion relation almost equal to the light line. Outside the light cone, the modes are nearly purely plasmonic, with a well-balanced mix only around the intersection.

In the far field, when all modes are radiating [Figs.~\ref{fig:Hopfield}(d) and \ref{fig:Hopfield}(f)], we observe an inversion in the predominance between photonic and plasmonic excitations at the crossing with a light line.
The band which radiates the most, i.e., the one located inside the light cone formed by light line $l$ that intersects the dispersion curve, is mainly plasmonic.
The other band, which in the near-field case corresponded to the nonradiating one, shows an equitable share between photonic and plasmonic excitations away from a light line.
When it approaches the intersection, it becomes predominantly photonic, as it can be inferred from the dispersion curves where this band falls into the light line.

Finally, we remark that for both polarizations, increasing the center-to-center distance leads to an increasingly equitable mix between photonic and plasmonic excitations, with both contribution tending to $0.5$. 
This is consistent with the fact that, as will be discussed in detail in the following subsection, when $d\gg a$, we end up with single noninteracting nanoparticles but which interact with the photonic continuum, leading to a polaritonic eigenstate with equal plasmonic and photonic weightings.

\subsection{Single nanoparticle limit}
\label{sec:Single NP Limit}

As mentioned above, we can use the limit of infinitely spaced nanoparticles, $d/a\!\to\!\infty$, to study the case of isolated single nanoparticles. These noninteracting nanoparticles are still coupled to the surrounding photonic continuum, leading to a radiative shift $\delta_0$ of the resonance frequency $\omega_0$, and to a radiative decay rate $\gamma_0$.

As discussed previously, when the nanoparticles are moved apart in the chain, the dispersion curves flatten and all the eigenfrequencies become degenerate at the eigenfrequency of a single nanoparticle.
All the quantities hence must become polarization and mode-independent and the analysis of the Hopfield coefficients, in the previous subsection, showed that one tends, indeed, toward a single degenerate mode of polaritonic nature, consisting of an equal mix of light and matter excitations.

The single nanoparticle case has already been studied perturbatively in Ref.~\cite{Downinga}, using a similar quantum formalism as in the present paper.
Using second-order perturbation theory (as was done for the chain, cf.\ Sec.~\ref{sec:Light Matter Interaction Perturbative}) the radiative shift $\delta_0$ is given by $\delta_0 = (E^{(2)}_{n^{\sigma}+1}  - E^{(2)}_{n^{\sigma}})/\hbar$, where the second-order energy correction reads
\begin{equation}
\label{eq:second order correction single NP}
    E^{(2)}_{n^{\sigma}} = \pi\hbar\omega_0^3\frac{a^3}{\mathcal{V}} \sum_{\mathbf{k},\hat\lambda_{\mathbf{k}}} \frac{(\hat\sigma \cdot \hat\lambda_{\mathbf{k}})^2}{\nu_{\mathbf{k}}} \frac{ (2n^{\sigma} +1)\nu_{\mathbf{k}} - \omega_0 }{\omega_0^2 - \nu_{\mathbf{k}}^2}.
\end{equation}
In the above expression, $n^{\sigma}$ is the quantum number corresponding to the number of plasmonic excitations with polarization $\sigma$, and $\nu_{\mathbf{k}} = \nu^{l=0}_{\mathbf{k}}$ is the photon dispersion, where, since the single nanoparticle is not anymore a periodic system, we do not have to take into account umklapp plasmon--photon processes. 

The correction \eqref{eq:second order correction single NP} appears to be linearly divergent. 
The authors of Ref.~\cite{Downinga} hence regularized such a divergence using a renormalization scheme analogous to that used by Bethe in his analysis of the Lamb shift in atomic physics \cite{Bethe}. 
Bethe's mass renormalization amounts to substract from Eq.~\eqref{eq:second order correction single NP} the energy shift corresponding to
free electrons coupled to the photonic continuum, which can be obtained taking the limit of vanishing transition frequency $\omega_0 \rightarrow 0$. 
Notably, this leads to a divergence which is only logarithmic.

However, as detailed in Sec.~\ref{sec:Light Matter Interaction Perturbative} in the case of the chain, the second-order correction is logarithmically divergent.
Trying to use the same renormalization procedure for the chain leads to unphysical results, with a singularity at the center of the Brillouin zone.
Furthermore, a perturbative treatment of the light--matter interaction in a similar plasmonic model but in two-dimensional lattices with arbitrary geometries leads to finite, nondivergent corrections, which agree well with classical electrodynamic calculations \cite{Fernique2020}.
This leads us to question the validity of Bethe's mass renormalization procedure in our specific case of a plasmonic system, for which the energy scales are quite different from the case of atomic physics studied originally by Bethe. In Appendix~\ref{sec:AppendixCommentDowning2017}, we also discuss the results obtained through this mass renormalization procedure for the case of nanoparticle dimers.

Keeping the linearly divergent second-order correction \eqref{eq:second order correction single NP}, one can go to the continuum limit  and transform the sum over wave vectors $\mathbf{k}$ into principal value integrals using spherical coordinates.
After calculating the integrals, the single nanoparticle radiative frequency shift reads
\begin{equation}
\label{eq:single NP radiative shift}
    \delta_0 = \frac{\omega_0}{3\pi}(k_0a)^3\left[\mathrm{ln}\left(\frac{\omega_\mathrm{c}/\omega_0+1}{\omega_\mathrm{c}/\omega_0-1} \right) - 2\,\frac{\omega_\mathrm{c}}{\omega_0} \right], 
\end{equation}
where, as expected, the expression now depends linearly on the cutoff $\omega_\mathrm{c}$ and hence also on the cutoff parameter $\Lambda$ (see footnote 3).

\begin{figure}[tb]
    \includegraphics[width=\columnwidth]{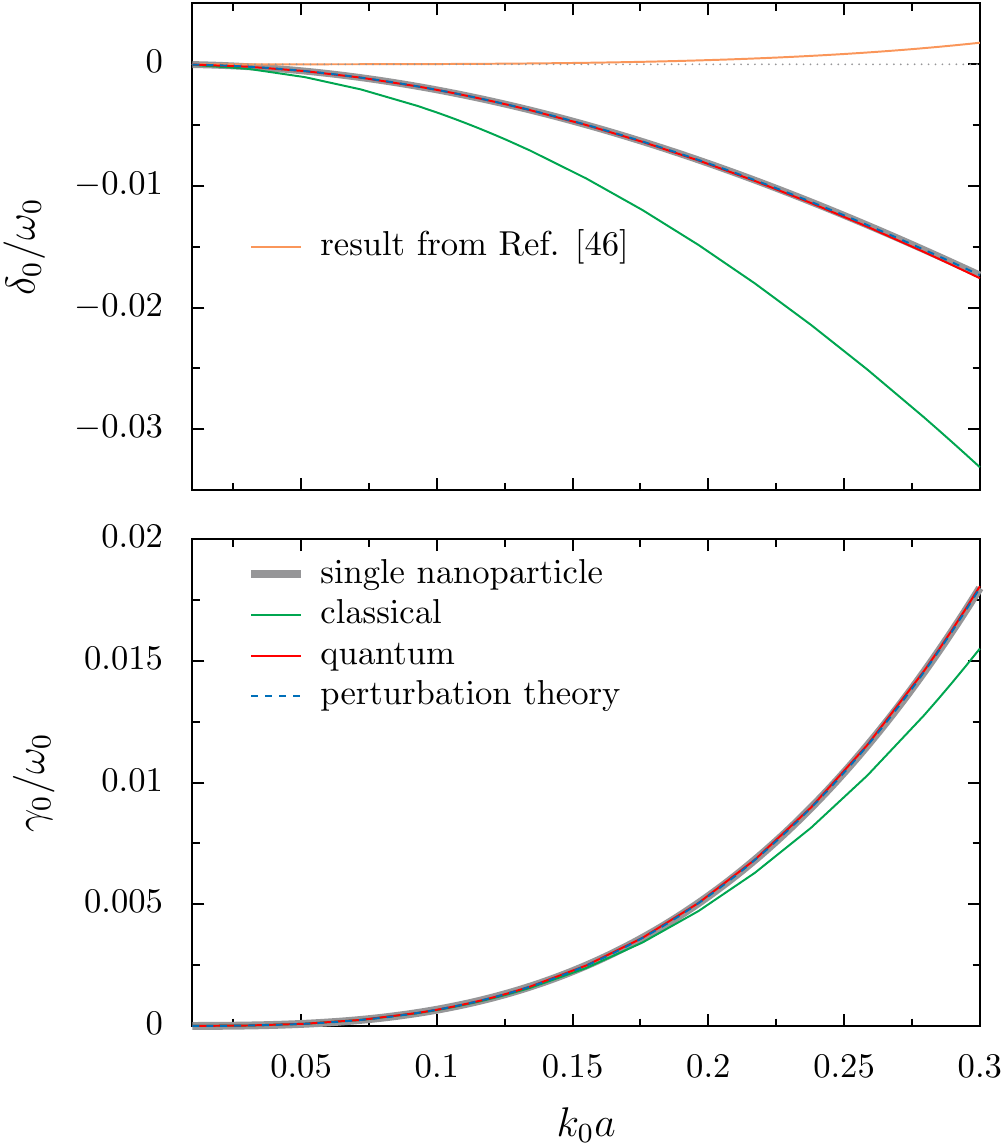}
    \caption{Single nanoparticle radiative shift $\delta_0$ (upper panel) and radiative decay rate $\gamma_0$ (lower panel) in units of the single nanoparticle resonance frequency $\omega_0$ and as a function of the reduced nanoparticle radius $k_0a$.
    The green solid line represents the result obtained from the classical model, as presented in Appendix~\ref{sec:AppendixClassicalModel}, while the thick grey solid line (almost perfectly overlapped by both the fully retarded quantum results in red and the perturbative results in dashed blue) represents the single nanoparticle results of Eqs.~\eqref{eq:single NP radiative shift} and \eqref{eq:single NP decay rate}.
    In the upper panel, the orange line represents the result for the radiative shift presented in Ref.~\cite{Downinga}, and the grey dotted line is a guide for the eye.}
\label{fig:LimitSingleNP}
\end{figure}

The upper panel of Fig.~\ref{fig:LimitSingleNP} shows the different results for the radiative shift $\delta_0$, according to the hypothesis used.
We see that the results from Ref.~\cite{Downinga} are significantly different both from those obtained by the classical model (green solid line) and by the quantum formalism where Bethe's mass renormalization procedure was not used (red solid line and blue dashed line) for the chain via the limit $d/a\!\to\!\infty$, and from the quantum model of a single nanoparticle via Eq.~\eqref{eq:single NP radiative shift}. 
Reference~\cite{Downinga} predicts a minute blueshift, while other methods predict a redshift which is one order of magnitude larger (in absolute value).
We see that, as expected, the results from the chain using our quantum formalism are in perfect agreement with the one from the single nanoparticle computation \eqref{eq:single NP radiative shift}, and the three curves are almost perfectly overlapping. 
The classical computation also leads to a redshift of the same order of magnitude but larger.
We note that in the classical formalism, using the radiative correction to the classical polarizability (which is defined in Appendix \ref{sec:AppendixClassicalModel}) leads to almost no frequency shift, while using the modified long wavelength approximation (MLWA, see Appendix~\ref{sec:AppendixClassicalModel}) leads to results which are almost the same as the one presented here for which we used the exact Mie polarizability.

The radiative decay rate $\gamma_0$ of the single nanoparticle can also be obtained perturbatively through our quantum formalism.
Using Fermi's golden rule similarly as for the chain in Sec.~\ref{sec:Light Matter Interaction Perturbative}, one obtains \cite{BrandstetterKunc}
\begin{equation}
\label{eq:single NP decay rate}
    \gamma_0 = \frac{2\omega_0^4a^3}{3c^3}.
\end{equation}
This familiar expression, representing the rate at which the plasmons dissipate their energy by emitting photons, can be obtained classically through the direct evaluation of the radiated power of an oscillating electric dipole \cite{Jackson2007}.

In the lower panel of Fig.~\ref{fig:LimitSingleNP}, we show the comparison between Eq.~\eqref{eq:single NP decay rate} and the results obtained by taking the limit of infinitely spaced nanoparticles in the chain.
We see that both the fully retarded and perturbative quantum results are in perfect agreement with the expression \eqref{eq:single NP decay rate}, as the three curves are almost indistinguishable. This indicates that taking the limit of infinite center-to-center distance $d$ is an accurate method to study the case of single nanoparticles.

We also observe in the lower panel of Fig.~\ref{fig:LimitSingleNP} that the classical chain gives slightly different results and predicts a smaller damping rate for values of $k_0a\gtrsim0.2$.
Notably, this difference depends on the approximation we adopt for the classical polarizability of the nanoparticle.
Indeed, using the radiative correction gives results which agree with Eq.~\eqref{eq:single NP decay rate}, while using the MLWA, which takes into account possible depolarization effects [see the discussion in Appendix \ref{sec:AppendixClassicalModel} after Eq.~\eqref{eq:Polarizability}] gives results similar to the one presented in Fig.~\ref{fig:LimitSingleNP} (see green solid line), where we used the Mie polarizability.\footnote{Note that the MLWA can be obtained from the Mie polarizability \eqref{eq:Polarizability} through an expansion in the small parameter $\omega a/c$.}
The small difference in the obtained results for the single nanoparticle radiative decay rate are thus due to depolarization effects, which increase with the (reduced) nanoparticle radius $k_0a$.
However, this cannot fully explain the small discrepancy in the decay rates of the chain discussed in Sec.~\ref{sec:Dispersion and decay rate}, as the classical approach slightly underestimates the rate compared to the quantum one, regardless of the approximation performed for the classical polarizability.

As can be seen from the lower panel of Fig.~\ref{fig:LimitSingleNP}, for a reduced radius $k_0a=0.3$, one obtains $\gamma_0=0.018\,\omega_0$. This means that in the case of the chain, several modes are superradiant. In particular, in the near-field case presented in Fig.~\ref{fig:dispersion}, all transverse radiating modes are superradiant, with a damping from $3$ to $9$ times larger than that of the single nanoparticle.
Interestingly, when we increase the center-to-center distance $d$ in Figs.~\ref{fig:dispersion_13a} and \ref{fig:dispersion_23a}, only the modes inside the light cone formed by the light line $l$ which intersects the dispersion curve (respectively, $l=+1$ and $l=-1$) present superradiance, while the ones outside this light cone are subradiant [see Figs.~\ref{fig:dispersion_13a}(c), \ref{fig:dispersion_13a}(d), \ref{fig:dispersion_23a}(c), and \ref{fig:dispersion_23a}(d)].


\section{Conclusion}
\label{sec:Conclusion}

We have provided a quantum theory of a linear chain of metallic nanoparticles hosting plasmon--polaritons. In particular, the retardation effects of the long-range dipole--dipole interaction along the chain have been considered exactly.
These retardation effects, arising from the light--matter interaction between the plasmonic degrees of freedom and a three-dimensional photonic continuum, lead to a polaritonic behavior of the eigenmodes of the system.
By taking into account plasmon--photon umklapp processes, we have been able to describe nanoparticles at arbitrary center-to-center distances $d\gtrsim3a$, that is, in both near- and far-field coupling regimes.

We have shown, by computing the band structure, the radiative linewidth, as well as the group velocities of the eigenmodes, that our quantum description is in good agreement with the usual classical approach used for plasmonic chains. 
The latter approach has also been shown to be in satisfying agreement with simulations using finite-difference time-domain techniques \cite{Pocock2018}. 
We have provided approximate analytical expressions for the above-mentioned quantities which reproduce the exact results, at least at a qualitative level, although they do not describe the polaritonic nature of the excitations.
These analytical expressions are valid in both near- and far-field situations and allow for a much more transparent formalism than in the classical case, where a numerical resolution is needed.

In the far-field regime and thanks to our analytical developments, we have observed a particularly unusual behavior of the group velocity, which changes signs and vanishes every time the center-to-center distance $d$ approximately equals a positive multiple of half of the resonance wavelength associated with the single isolated nanoparticle. The group velocity of any polaritonic mode can hence be tuned to a variety of different values, including zero and large positive or negative group velocities, depending on the distance $d$.

We have studied quantitatively the hybridization of plasmonic and photonic degrees of freedom using Hopfiled's coefficients, showing a well-balanced hybridization inside the light cone even in the longitudinal polarization case, which leads us to associate the presence of plasmon--polariton excitations to the presence of radiative losses rather than to the presence of an avoided crossing dispersion.

Finally, we have studied the regime of infinitely spaced nanoparticles and found eigenfrequencies and radiative decay rates in perfect agreement with the single nanoparticle case. Notably, we have found results that are qualitatively different from the ones of Ref.~\cite{Downinga}. 
We showed that these differences arise from a mass renormalization scheme that may not be adequate to the present situation and that we did not follow here.

Our detailed analysis supports the development of a quantum formalism, which is required to have a self-contained theory of arrays of interacting nanoparticles, where quantum effects, such as Landau damping or electronic spill-out, can be readily incorporated.
Our work also opens the perspective of studying other systems where light--matter interaction is of major importance, such as cavity-embedded systems, arrays of larger dimensionality and/or with other geometries, and plasmonic systems composed of nonspherical nanoparticles.


\begin{acknowledgments}
We thank Charles A.\ Downing, Charlie-Ray Mann, and Eros Mariani for insightful discussions.
We acknowledge financial support from the French National Research Agency (ANR) through Grant No.\ ANR-14-CE26-0005 Q-MetaMat. 
This work of the Interdisciplinary Thematic Institute QMat, as part of the ITI 2021-2028 program of the University of Strasbourg, CNRS, and Inserm, was supported by IdEx Unistra (ANR 10 IDEX 0002), and by SFRI STRAT’US Projects No.\ ANR-20-SFRI-0012 and No.\ ANR-17-EURE-0024 under the framework of the French Investments for the Future Program.
\end{acknowledgments}

\appendix

\section{Classical electromagnetic modeling of the nanoparticle chain}
\label{sec:AppendixClassicalModel}

The standard theoretical approach for describing the plasmonic chain introduced in the main text (see Fig.~\ref{fig:sketch}) is based on macroscopic classical electrodynamics and benefits from extensive literature \cite{Quinten1998,Park,Weber,Koenderink,Citrin2004,Citrin2006,Fung2007, Markel, Conforti2010,Udagedara,Rolly2012}, which is still active nowadays \cite{Compaijen,Pikalov}. 
In this appendix, we describe the dielectric properties of the metallic nanoparticles using the Drude model. Neglecting Ohmic losses, the nanoparticle local permittivity is given by $\epsilon(\omega) = 1 - \omega^2_{\mathrm{p}}/\omega^2$. 

We first start by looking for the polarizability of a single nanoparticle. Using Mie's theory \cite{Mie,Bohren2004} and considering only the electric dipole term, one obtains
\begin{align}
    \label{eq:Scattered field}
    \mathbf{E_\mathrm{s}} =&\, \frac{3\mathrm{i}c^3E_0a_1(\omega)}{2\omega^3}\frac{\mathrm{e}^{\mathrm{i}\omega r/c}}{r}
    \bigg \{ \frac{\omega^2}{c^2}({\hat{r}}\times{\hat{u}})\times{\hat{r}} \nonumber\\
    &\,+ [ 3{\hat{r}}({\hat{r}}\cdot{\hat{u}}) - {\hat{u}} ]\left(\frac{1}{r^2} - \frac{\mathrm{i}\omega}{cr}\right) \bigg \}
\end{align}
for the expression of the electric field scattered by the sphere.
Here, $E_0$ is the amplitude of the incident field $\mathbf{E}_{\mathrm{i}}=E_{0}\mathrm{e}^{\mathrm{i}\omega r\cos\theta/c}{\hat{u}}$ in a given direction $\hat{u}$, and
\begin{equation}
\label{eq:Mie_coeff}
   a_1(\omega)=\frac{n\psi_1\left(nka\right)\psi'_1\left(ka\right) -  \psi_1\left(ka\right)\psi'_1\left(nka\right)}{n\psi_1\left(nka\right)\xi'_1\left(ka\right) -  \xi_1\left(ka\right)\psi'_1\left(nka\right)}
\end{equation}
is the first (electric dipole) Mie coefficient \cite{Capolino2009}, where $n = \sqrt{\epsilon(\omega)}$ is the refractive index, $k=\omega/c$ the wave number in the metal, and $a$ the nanoparticle radius.
$\psi_1(x) = \sin(x)/x - \cos(x)$ and $\xi_1(x)=(-\mathrm{i}/x - 1)\mathrm{e}^{\mathrm{i}x}$ are, respectively, the first Ricatti--Bessel functions of the first and second kinds.
The scattered field \eqref{eq:Scattered field} corresponds to the radiation field emitted by an oscillating electric dipole, therefore taking fully into account retardation effects, with a dipolar moment $\mathbf{p}=(3\mathrm{i}/2)(c/\omega)^3E_0a_1(\omega){\hat{u}}$.
Using that the dipolar moment is also equal to $\mathbf{p}=\alpha(\omega)\mathbf{E_i}|_{\substack{r=0}}$ leads to an expression of the dynamic, frequency-dependent polarizability of the metallic nanoparticle \cite{Doyle,Fung2007,Rolly2012},
\begin{equation}
    \label{eq:Polarizability}
    \alpha(\omega)=\left(-\mathrm{i}\frac{2\omega^3}{3c^3}\right)^{-1}a_1(\omega).
\end{equation}

Many authors \cite{Weber,Koenderink,Pikalov,Udagedara,Markel} use an \textit{ad hoc} correction to the quasistatic polarizability of a spherical nanoparticle
\begin{equation}
    \alpha_{\mathrm{qs}}(\omega) = a^3\frac{\epsilon(\omega) -1}{\epsilon(\omega)+2},
\end{equation}
instead of the exact Mie expression \eqref{eq:Polarizability}. To take into account the radiative decay of the plasmons and preserve energy conservation, the choice $\alpha^{-1}_{\mathrm{ad hoc}} = \alpha^{-1}_{\mathrm{qs}} -\mathrm{i}2\omega^3/3c^3$ is usually made.
Such a radiative correction can be found by adding a radiation reaction field to the quasistatic system \cite{Wokaun1982,Draine,Jackson2007}. 
Other authors \cite{Compaijen} also include a contribution proportional to $\omega^2$ to the inverse polarizability, stemming from the introduction of a depolarization field in the system. 
This is known as the modified long wavelength approximation (MLWA), and it leads to a redshift of the polaritonic dispersion curves which has been interpreted as arising from a dephasing between radiation emitted by different parts of the sphere~\cite{Moroz}. 

As already discussed in Refs.~\cite{Meier1983,Capolino2009,Grigoriev}, an expansion of the exact polarizability \eqref{eq:Polarizability} for small particles to third order in $\omega a/c$ leads, up to a numerical factor, to the same \textit{ad hoc} corrections mentioned above.

Since the radiative correction and the MLWA are approximations of Eq.~\eqref{eq:Polarizability}, and since using the latter equation does not increase substantially the computation time, in this paper we have chosen to describe the nanoparticles within the classical model using the exact Mie polarizability \eqref{eq:Polarizability}.
Following Ref.~\cite{Koenderink}, one can then obtain an implicit dispersion relation by writing the induced dipole moment $\mathbf{p}_n$ of the nanoparticle $n$ as $\mathbf{p}_n=\alpha(\omega)\mathbf{E}_{\mathrm{loc}}$, where $\mathbf{E}_{\mathrm{loc}}$ is the total electric field generated by all the others nanoparticles $m\neq n$. One finds
\begin{equation}
    \label{eq:Classical dispersion}
    1 + \frac{\alpha(\omega)}{d^3}\Sigma_{\sigma}(\omega, q) = 0,
\end{equation}
where $q$ is the wave number of the normal mode along the chain, belonging to the first Brillouin zone. 
Equation~\eqref{eq:Classical dispersion} can be seen as the classical analog of Eq.~\eqref{eq:Eigenfreq_exact_developed}.
As it has been shown in Ref.~\cite{Citrin2006}, the function $\Sigma_{\sigma}(\omega,q)$ can be rewritten, using analytical continuation, in terms of a sum of  polylogarithms,
\begin{subequations}
    \begin{align}
        \Sigma_{z}(\omega, q) =&\, 2\mathrm{i}\frac{\omega d}{c}\left[ \mathrm{Li}_2(\varphi^+) + \mathrm{Li}_2(\varphi^-)\right] \nonumber\\
        &\,- 2\left[\mathrm{Li}_3(\varphi^+) + \mathrm{Li}_3(\varphi^-) \right], \\
        \Sigma_{x,y}(\omega, q) =&\, -\left(\frac{\omega d}{c}\right)^2\left[ \mathrm{Li}_1(\varphi^+) + \mathrm{Li}_1(\varphi^-)\right] \nonumber\\
        &\,- \mathrm{i}\frac{\omega d}{c}\left[\mathrm{Li}_2(\varphi^+) + \mathrm{Li}_2(\varphi^-) \right]  \nonumber\\
        &\,+ \left[ \mathrm{Li}_3(\varphi^+) + \mathrm{Li}_3(\varphi^-)\right],
    \end{align}
\end{subequations}
where $\varphi^{\pm} = \mathrm{e}^{\mathrm{i}(\omega/c \pm q)d}$.
Here, $\sigma=x$, $y$ or $z$ characterizes the polarization of the collective excitation, depending on whether the induced dipole moment points along the array ($\sigma=z$) or orthogonal to it ($\sigma=x,y$).
Since collective and individual LSP polarizations are aligned, it is equivalent to consider a given polarization $\sigma$ on each individual LSP.

Numerically solving Eq.~\eqref{eq:Classical dispersion} for complex $\omega$ leads to the dispersion and to the radiative linewidth of the polaritonic normal modes discussed in the main text (see green solid lines in 
Figs.~\ref{fig:dispersion}--\ref{fig:dispersion_23a}).

\section{The diamagnetic $A^2$ term}
\label{sec:AppendixDiamagneticTerm}

In our treatment of the photonic environment, we discarded the $A^2$ term, also known as the diamagnetic term. In this appendix, we discuss the origin of such a term and give justifications for this approximation.

Physically, the diamagnetic term represents a photon self-interaction energy which is not involved in the retardation effects of the light--matter coupling. It comes from the minimal-coupling substitution of the momentum $\mathbf{P} \xrightarrow{} \mathbf{P} + (e/c)\mathbf{A}(\mathbf{r})$ in the kinetic Hamiltonian $H_{\mathrm{kin}} = \mathbf{P}^2/2m$, with $m$ the considered mass.
We recall that this coupling originates fundamentally in a relativistic theory by the replacement in the free Lagrangian of the usual derivatives with the covariant derivatives,  $\partial_{\mu} \xrightarrow{} D_{\mu} = \partial_{\mu} - (\mathrm{i}e/c)A_{\mu}$, where $A_{\mu}$ is the four-vector gauge potential. 
This replacement is required to preserve the $U(1)$ gauge symmetry of electromagnetism, therefore, the light--matter coupling is fully determined by the gauge invariance of the theory itself \cite{Schwartz}.

Neglecting the diamagnetic term can thus lead to conceptual problems such as the loss of gauge invariance but also unphysical ground states \cite{Rokaj2018,Schaefer2020}. It is, however, frequently discarded due to its quadratic nature in the light--matter coupling and several well-known quantum optics models such as the Jaynes--Cummings or Dicke model do not consider it. 
Perceptible differences due to the diamagnetic term such as gauge ambiguities were found in the context of the ultra strong coupling (USC) or deep strong coupling (DSC) regime only \cite{Stefano2019}. 

Including the $A^2$-term in our model amounts to replacing Eq.~\eqref{eq:H_plph} by the quadratic plasmon--photon coupling Hamiltonian
\begin{equation}
\label{eq:H_plphA^2}
 H^{\sigma, A^2}_{\mathrm{pl}\textrm{-}\mathrm{ph}} = \frac{e}{m_{\mathrm{e}}c} \sum_{n=1}^{\mathcal{N}} \mathbf{\Pi}^{\sigma}_n \cdot \mathbf{A} (\mathbf{d}_n) +  \frac{N_{\mathrm{e}} e^2}{2 m_{\mathrm{e}}c^2} \sum_{n=1}^{\mathcal{N}} \mathbf{A}^2 (\mathbf{d}_n).
\end{equation}
Since the quadratic $A^2$ term does not involve plasmonic degrees of freedom, it has no effect when one applies perturbation theory on the plasmonic Hamiltonian \eqref{eq:H_pl}, except for a global energy shift \cite{Downing2017}. Therefore, the results found through the perturbative treatment in Sec.~\ref{sec:Light Matter Interaction Perturbative} are the same, whether we include such a term or not.
However, it can renormalize the photonic dispersion by inducing what is usually referred to as a diamagnetic shift, hence slightly modifying the polaritonic dispersion.

To gain insight into the effects of such a term in our system, one can diagonalize the Hopfield model, corresponding to a model close to the one studied here but with a three-dimensional dipolar system instead of a one-dimensional one, with and without the quadratic term and show that different behaviors occur only for the USC and DSC regimes \cite{Kockum2019_review}. 
We recall that one enters in the USC regime when the counter-rotating terms of the Hamiltonian become sizable \cite{FornDiaz2019}. From the analysis of the Hopfield coefficients in Sec.~\ref{sec:Hopfield}, one can then deduce that in the model studied here, we are far from the USC regime and therefore that the diamagnetic term can be safely neglected.

\section{Analytical expressions for the Hopfield coefficients}
\label{sec:AppendixHopfield}
In Sec.~\ref{sec:Hopfield}, we discussed the behavior of the Hopfield coefficients, namely the plasmonic and photonic weightings of the polaritonic excitation.
Here, we provide the analytical expressions for the latter weightings, which are displayed in Fig.~\ref{fig:Hopfield}.
Using the system of equations \eqref{eq:system of equations}, one can write the photonic part of the weightings of the excitation as
\begin{equation}
    \sum_l \sum_{\boldsymbol{\kappa}, \hat{\lambda}^l_{\boldsymbol{\kappa}q}} \left| Y_{\boldsymbol{\kappa}q}^{l,\sigma,\hat{\lambda}^l_{\boldsymbol{\kappa}q}} \right|^2 = \frac{4\omega_0^2{\omega_q^{\sigma}}^4\left|X_q^{\sigma}\right|^2}{\left|{\omega_q^{\sigma}}^2 - \omega_0\Omega_q^{\sigma}\right|^2} \sum_l \mathcal{I}_{q,-}^{l,\sigma}
\end{equation}
and 
\begin{equation}
    \sum_l \sum_{\boldsymbol{\kappa}, \hat{\lambda}^l_{\boldsymbol{\kappa}q}} \left| Z_{\boldsymbol{\kappa}q}^{l,\sigma,\hat{\lambda}^l_{\boldsymbol{\kappa}q}} \right|^2 = \frac{4\omega_0^2{\omega_q^{\sigma}}^4\left|X_q^{\sigma}\right|^2}{\left|{\omega_q^{\sigma}}^2 - \omega_0\Omega_q^{\sigma}\right|^2} \sum_l \mathcal{I}_{q,+}^{l,\sigma},
\end{equation}
and the plasmonic weights as
\begin{equation}
    \left|W_q^{\sigma}\right|^2 = \left|X_q^{\sigma}\right|^2\left|\frac{{\omega_q^{\sigma}}^2 + \omega_0\Omega_q^{\sigma}}{{\omega_q^{\sigma}}^2 - \omega_0\Omega_q^{\sigma}}\right|^2
\end{equation}
and
\begin{align}
    \left|X_q^{\sigma}\right|^2 =&\, \Bigg( \frac{4\omega_0^2{\omega_q^{\sigma}}^4}{\left|{\omega_q^{\sigma}}^2 - \omega_0\Omega_q^{\sigma}\right|^2}\sum_l\left[ \mathcal{I}_{q,-}^{l,\sigma} - \mathcal{I}_{q,+}^{l,\sigma}  \right]   \nonumber \\
    &\, + \left|\frac{{\omega_q^{\sigma}}^2 + \omega_0\Omega_q^{\sigma}}{{\omega_q^{\sigma}}^2 - \omega_0\Omega_q^{\sigma}}\right|^2 - 1 \Bigg)^{-1}.
\end{align}
In the above expressions, $\mathcal{I}_{q,\pm}^{l,\sigma}$ are functions of the complex eigenfrequencies $\Omega_q^{\sigma}$, and read
\begin{equation}
    \mathcal{I}_{q,\pm}^{l,\sigma} = \sum_{\boldsymbol{\kappa}, \hat{\lambda}^l_{\boldsymbol{\kappa}q}} \frac{ \left(\xi_{\boldsymbol{\kappa}q}^l\right)^2 (\hat{\sigma} \cdot \hat{\lambda}_{\boldsymbol{\kappa}q}^l)^2}{\left|\Omega_q^{\sigma} \pm \nu_{\boldsymbol{\kappa}q}^l\right|^2}.
\end{equation}
Summing over the photon polarizations using Eq.~\eqref{eq:polarization sum}, and transforming the sum over $\boldsymbol{\kappa}$ to integrals via Eq.~\eqref{eq:continuum limit}, leads to the following result:
\begin{align}
     \mathcal{I}_{q,\pm}^{l,\sigma} =&\, \pm\frac{\eta^\sigma\omega_0a^3}{4dc^2}\left|\frac{cq^l}{\Omega_q^\sigma}\right|^2\Theta(\omega_\mathrm{c} - c|q^l|)\Bigg\{ \mp\left( \frac{1}{\omega_\mathrm{c}} - \frac{1}{c|q^l|} \right) \nonumber  \\
     &\, - \frac{\Omega_q^{\mathrm{r},\sigma}}{|\Omega_q^\sigma|^2}\ln\left( \frac{(\Omega_q^{\mathrm{i},\sigma})^2 + (\Omega_q^{\mathrm{r},\sigma} \pm c|q^l|)^2}{(\Omega_q^{\mathrm{i},\sigma})^2 + (\Omega_q^{\mathrm{r},\sigma} \pm \omega_\mathrm{c})^2}\frac{\omega_\mathrm{c}^2}{(c|q^l|)^2} \right) \nonumber \\
     &\,  + \frac{ \tan^{-1}\left( \frac{\Omega_q^{\mathrm{r},\sigma} \pm \omega_\mathrm{c}}{\Omega_q^{\mathrm{i},\sigma}} \right) - \tan^{-1}\left( \frac{\Omega_q^{\mathrm{r},\sigma} \pm c|q^l|}{\Omega_q^{\mathrm{i},\sigma}} \right)     }{\Omega_q^{\mathrm{i},\sigma}} \nonumber \\
     &\,\times \left[  \mathrm{sgn}\{\eta^\sigma\}\left|\frac{\Omega_q^\sigma}{cq^l}\right|^2 + \frac{(\Omega_q^{\mathrm{r},\sigma})^2 - (\Omega_q^{\mathrm{i},\sigma})^2}{|\Omega_q^\sigma|^2}   \right]     \Bigg\},
\end{align}
where $q^l = q - 2\pi l/d$, and where, for clarity, we have written the real and imaginary parts of the polaritonic eigenfrequencies $\Omega_q^\sigma$ as $\Omega_q^{\mathrm{r},\sigma}$ and $\Omega_q^{\mathrm{i},\sigma}$, respectively.
We note that the term in the third line of the above equation is well-defined due to the fact that the imaginary part $\Omega_q^{\mathrm{i}, \sigma} \rightarrow 0$ only when the real part $\Omega_q^{\mathrm{r}, \sigma} < c|q^l|$, i.e., guided modes which are immune to radiation damping exist only outside the first light cone, as discussed in Sec.~\ref{sec:Results}.

\section{Discussion of Bethe's mass renormalization procedure for nanoparticle dimers}
\label{sec:AppendixCommentDowning2017}
As shown in Sec.~\ref{sec:Single NP Limit}, we obtain drastically different results when we do not apply Bethe's mass renormalization procedure to our quantum model of a single metallic nanoparticle, and this led us to question the use of Bethe's approach in the specific case of a plasmonic system.
The radiative shift induced by the vacuum electromagnetic modes on a dimer of interacting metallic nanoparticles was computed in Ref.~\cite{Downinga} using Bethe's scheme.
In this appendix, we will reproduce the same quantities as presented in Ref.~\cite{Downinga}, but without using the mass renormalization procedure, and comment on the differences obtained.

The dimer of interacting metallic nanoparticles is modeled by the plasmonic Hamiltonian of the chain \eqref{eq:H_pl} for which we set the number of nanoparticles $\mathcal{N}=2$.
The dimer interacts with a photonic environment, and since the system is not periodic, we do not have to take into account umklapp plasmon--photon processes. 
The Hamiltonian describing the photonic environment is thus given by Eq.~\eqref{eq:H_ph} for which we only keep the term with $l=0$.
The coupling Hamiltonian is therefore given by Eq.~\eqref{eq:H_plph} with $\mathcal{N}=2$ and $l=0$.

As detailed in Ref.~\cite{Downinga}, such a system hosts bright and dark hybridized collective plasmonic modes, and one can easily diagonalize the plasmonic Hamiltonian to find the bare (quasistatic) eigenfrequencies,
\begin{equation}
    \omega_\tau^\sigma = \omega_0\sqrt{1+2\tau|\eta^\sigma|\frac{\Omega}{\omega_0}}.
\end{equation}
The label $\tau=\pm$ distinguishes the high- and low-energy coupled plasmonic modes. Importantly, the high-energy transverse ($\uparrow\uparrow$) and low-energy longitudinal ($\rightarrow\rightarrow$) excitations correspond to symmetric bright modes coupled to the photonic continuum, while the low-energy transverse ($\uparrow\downarrow$) and high-energy longitudinal ($\rightarrow\leftarrow$) excitations are antisymmetric dark modes, hence weakly coupled to the electromagnetic continuum.
Just as in the single-nanoparticle case or for the chain, these plasmonic modes will hybridize with the electromagnetic continuum, leading to dressed eigenfrequencies $\tilde\omega_\tau^\sigma = \omega_\tau^\sigma + \delta_\tau^\sigma$, where $\delta_\tau^\sigma$ are the radiative frequency shifts.

To compute the dressed eigenfrequencies, we rely on Ref.~\cite{Downinga}, which uses a similar methodology as the one we detailed in Sec.~\ref{sec:Light Matter Interaction Perturbative}. 
Without making use of the mass renormalization procedure, we obtain $\delta_\tau^{\sigma} = (E^{(2)}_{n_\tau^{\sigma}+1}  - E^{(2)}_{n_\tau^{\sigma}})/\hbar$, where the second-order correction reads
\begin{align}
    E^{(2)}_{n_\tau^{\sigma}} =&\, \pi\hbar\omega_0^2\omega_{\tau}^\sigma\frac{a^3}{\mathcal{V}}\sum_{\mathbf{k},\hat\lambda_{\mathbf{k}}} \frac{(\hat\sigma \cdot \hat\lambda_{\mathbf{k}})^2}{\nu_{\mathbf{k}}} \frac{ (2n^\sigma_\tau +1)\nu_{\mathbf{k}} - \omega_\tau^{\sigma}}{{\omega_\tau^\sigma}^2 - {\nu_{\mathbf{k}}}^2} \nonumber \\
    &\, \times\left[ 1 + \tau\,\mathrm{sgn}\{\eta^\sigma\}\cos\left(k_zd\right)\right].
\end{align}
Computing the summation over photon polarizations through the relation \eqref{eq:polarization sum} and transforming the sum over wave vectors $\mathbf{k}$ into a principal value integral in spherical coordinates, we get
\begin{align}
    \delta_\tau^\sigma =&\, 2\pi\omega_0^2\omega_\tau^\sigma\frac{a^3}{8\pi^3}\mathcal{P}\int_0^{k_\mathrm{c}}\frac{k^2\mathrm{d}k}{{{\omega_\tau^\sigma}^2 - c^2k^2}} \nonumber \\
    &\,\times \int_0^\pi \mathrm{d}\theta\sin\theta\left[1 + \tau\,\mathrm{sgn}\{\eta^\sigma\}\cos\left(kd\cos\theta\right)\right] \nonumber \\
    &\,\times \int_0^{2\pi} \mathrm{d}\varphi\left[1-(\hat\sigma \cdot \hat k)^2\right].
\end{align}
After a long but straightforward calculation, we obtain the radiative shifts
\begin{align}
\label{eq:shift dimer}
    \delta_\tau^\sigma =&\, \frac{{\omega_\tau^\sigma}^2}{3\pi\omega_0} \left(k_0a\right)^3 \left[\mathrm{ln}\left(\frac{\omega_\mathrm{c}/\omega_\tau^\sigma+1}{\omega_\mathrm{c}/\omega_\tau^\sigma-1} \right) - 2\frac{\omega_\mathrm{c}}{\omega_\tau^\sigma} \right] \nonumber \\
    &\, + \tau|\eta^\sigma|\frac{\omega_\tau^\sigma\Omega}{\pi\omega_0}\left(k_0d\right)^2 \bigg[ \left( \frac{1+\mathrm{sgn}\{\eta^\sigma\}}{2} \right)g_\tau^\sigma \nonumber \\
    &\, + \frac{h_\tau^\sigma}{k_\tau^\sigma d} - \frac{2\mathrm{Si}(k_\mathrm{c}d)+g_\tau^\sigma}{\left(k_\tau^\sigma d\right)^2}\bigg],
\end{align}
where $k_\tau^\sigma=\omega_\tau^\sigma/c$,
\begin{align}
    g_\tau^\sigma =&\, \sum_{\zeta=\pm}\bigg[ \zeta\sin(k_\tau^\sigma d)\mathrm{Ci}(k_\mathrm{c}d+\zeta k_\tau^\sigma d) \nonumber \\
    &\, - \cos(k_\tau^\sigma d)\mathrm{Si}(k_\mathrm{c}d+\zeta k_\tau^\sigma d) \bigg],
\end{align}
\begin{align}
    h_\tau^\sigma =&\, \sum_{\zeta=\pm}\bigg[ \zeta\cos(k_\tau^\sigma d)\mathrm{Ci}(k_\mathrm{c}d+\zeta k_\tau^\sigma d) \nonumber \\
    &\, + \sin(k_\tau^\sigma d)\mathrm{Si}(k_\mathrm{c}d+\zeta k_\tau^\sigma d) \bigg],
\end{align}
and with $\mathrm{Si}(x)$ and $\mathrm{Ci}(x)$ denoting the sine and cosine integrals, respectively. As expected and as for the single nanoparticle case [see Eq.~\eqref{eq:single NP radiative shift}], the expression \eqref{eq:shift dimer} depends linearly on the cutoff $\omega_\mathrm{c}$.

\begin{figure}[tb]
    \includegraphics[width=\columnwidth]{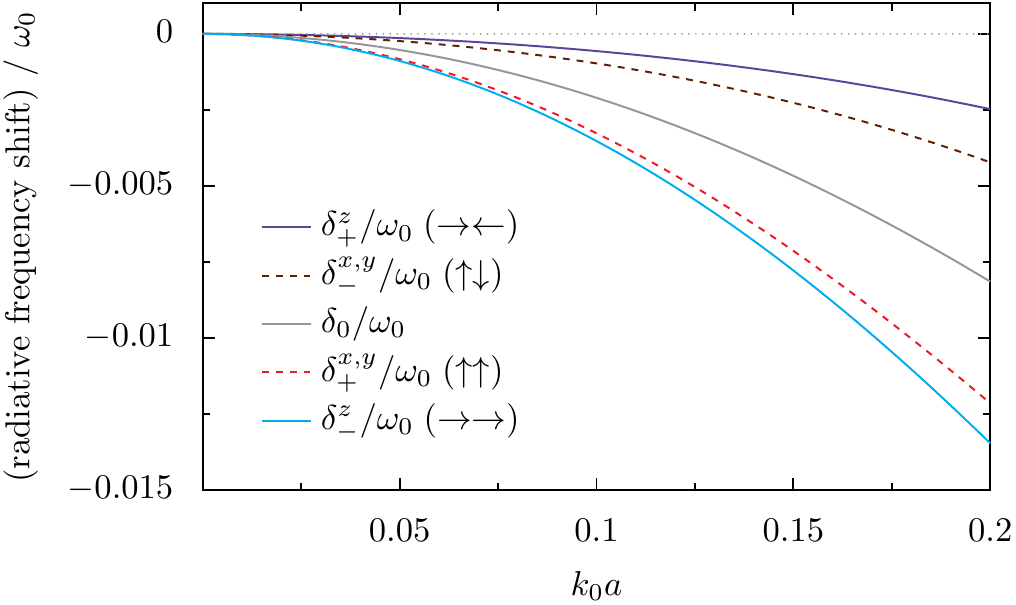}
    \caption{Radiative frequency shifts in units of the bare frequency $\omega_0$ as a function of the reduced nanoparticle radius $k_0a$. The same approach as for the chain and two-dimensional systems has been used, i.e., without making use of Bethe's mass renormalization. The grey dotted line is a guide for the eye. The interparticle distance is $d=3a$.}
\label{fig:DimersShift}
\end{figure}

In Fig.~\ref{fig:DimersShift}, we show the resulting radiative shifts for the dimer, together with the shift \eqref{eq:single NP radiative shift} obtained for the single nanoparticle. 
In Ref.~\cite{Downinga}, a blueshift has been obtained for the bright modes, whereas only the dark modes were redshifted. This resulted in an increase (decrease) of the splitting between hybridized plasmonic modes for the transverse
(longitudinal) polarization.
Here, we see that both the bright and dark modes are redshifted, coherently with the radiative redshifts observed for the chain and also for two-dimensional lattices \cite{Fernique2020}.
The bright modes are significantly more altered than the dark ones, which is consistent with the fact that they interact more strongly with light.
Therefore, the splittings we predict are reversed from those presented in Ref.~\cite{Downinga}, since here we show an increase (decrease) of the splitting between hybridized plasmonic modes for the longitudinal (transverse) polarization.
Importantly, the frequency shifts we predict are also more than one order of magnitude larger than the ones presented in Ref.~\cite{Downinga}.
Hence, the experimental protocol proposed in Ref.~\cite{Downinga} to detect the frequency splitting between bright and dark modes could be more accessible than originally thought.

Another quantity that has been proposed to be observed experimentally is the ratio $\Delta\tilde\omega^z/\Delta\tilde\omega^{x,y}$ of the longitudinal and transverse frequency splittings $\Delta\tilde\omega^\sigma = \tilde\omega^\sigma_+ - \tilde\omega^\sigma_-$ between bright and dark modes.
In the absence of light--matter interaction, the bare ratio $\Delta\omega^z/\Delta\omega^{x,y}=2$ is independent of the center-to-center distance $d$ up to quadratic corrections in $\Omega/\omega_0\ll1$.
In Ref.~\cite{Downinga}, it was found that the ratio $\Delta\tilde\omega^z/\Delta\tilde\omega^{x,y}$ presents a universal scaling with the distance $d$, being independent of the nanoparticle radius $a$, unlike the radiative frequency shifts for individual dimer levels.

\begin{figure}[tb]
    \includegraphics[width=.92\columnwidth]{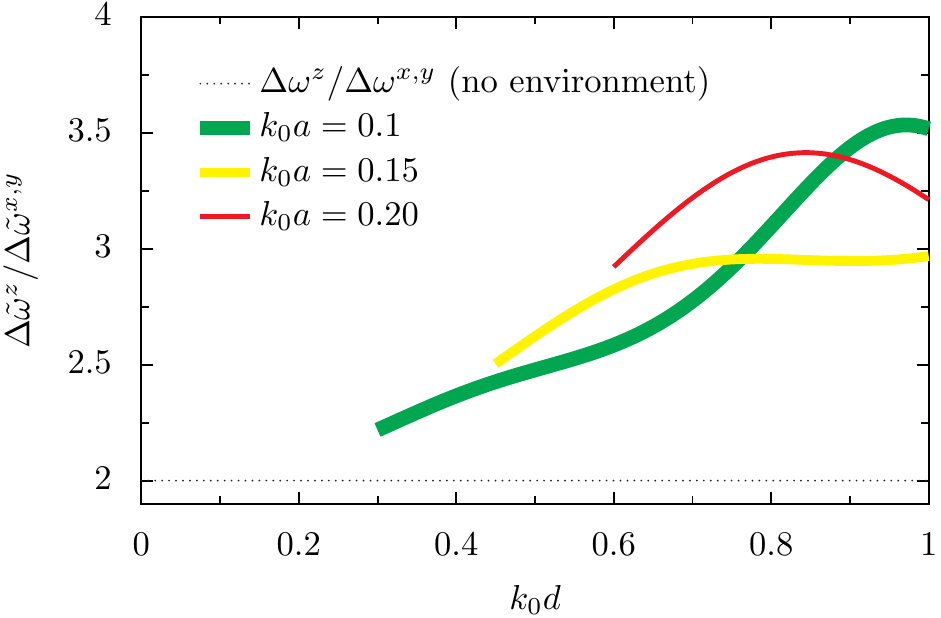}
    \caption{Ratio $\Delta\tilde\omega^z/\Delta\tilde\omega^{x,y}$ of longitudinal and transverse frequency splittings in a nanoparticle dimer as a function of the (reduced) interparticle distance $k_0d$, for increasing (reduced) nanoparticle radius $k_0a$, from $0.1$ (green line), $0.15$ (yellow line), to $0.2$ (red line). The grey dotted line shows the bare ratio $\Delta\omega^z/\Delta\omega^{x,y}$ in the absence of the coupling with the photonic environment. The UV frequency cutoff is chosen as $\omega_\mathrm{c}=c/a$.}
\label{fig:DimersDelta}
\end{figure}

In Fig.~\ref{fig:DimersDelta}, we show the result obtained for the dimensionless ratio $\Delta\tilde\omega^z/\Delta\tilde\omega^{x,y}$ using our approach that does not use Bethe's mass renormalization.
We immediately see that the universal scaling put forward in Ref.~\cite{Downinga} is no longer present, implying that the dimensionless ratio $\Delta\tilde\omega^z/\Delta\tilde\omega^{x,y}$ depends on the nanoparticle radius.
Actually, this makes the experimental realization of the protocol proposed in Ref.~\cite{Downinga} to detect the level splitting and this dimensionless ratio even more interesting, since it could allow one to gauge the relevance of Bethe's renormalization procedure for dipolar systems such as the one studied in this appendix.

To conclude, this appendix demonstrated that not using Bethe's mass renormalization procedure leads to drastically different conclusions about the frequency shifts induced by the light--matter interaction in metallic nanoparticle dimers.
Here, we predict radiative frequency shifts of more than one order of magnitude larger than the one predicted in Ref.~\cite{Downinga}, an increase of the splitting between longitudinal bright and dark modes, and a dependence of the dimensionless ratio $\Delta\tilde\omega^z/\Delta\tilde\omega^{x,y}$ on nanoparticle radius.
Importantly, within this approach, both the single nanoparticle and dimer radiative frequency shifts depend linearly on the choice of cutoff $\omega_\mathrm{c}$, as opposed to the chain radiative frequency shift which depends on it only logarithmically, and to two-dimensional lattices radiative shifts which are cutoff independent \cite{Fernique2020}. In the three-dimensional case, an exact diagonalization which does not require any cutoff is even possible \cite{Lamowski}. This suggests that our quantum formalism of light--matter-induced radiative frequency shift may be more adapted to the study of periodic systems such as lattices rather than to single nanoparticles or dimers.


\bibliography{report_jabref}

\end{document}